\documentclass[aps,floatfix,a4paper,twoside,twocolumn]{revtex4}
\usepackage{graphicx,amssymb,amsmath,bbm,mathrsfs,times}
\usepackage{pstricks}
\newcommand{\ket}[1]{\vert #1 \rangle}
\newcommand{\bra}[1]{\langle #1 \vert}
\newcommand{\dket}[1]{\vert #1 \rangle\rangle}
\newcommand{\dbra}[1]{\langle\langle #1 \vert}

\newcommand{\media}[1]{\langle #1 \rangle}
\newcommand{\calpha}{\alpha^*} \newcommand{\cbeta}{\beta^*}
\newcommand{\cxi}{\xi^*} \newcommand{\czeta}{\zeta^*}
\newcommand{\cv}{v^*}  \newcommand{\cw}{w^*}
\newcommand{\iid}{\mathbb{I}}  \newcommand{\wtB}{B}
\newcommand{\bmsigma}{\boldsymbol \sigma} \newcommand{\bmX}{\boldsymbol X}
\newcommand{\bbGamma}{{\rm I}\!\Gamma}

\newcommand{\witA}{\widetilde{A}} \newcommand{\witB}{\widetilde{B}}
\begin{document}
\title{De-Gaussification by inconclusive photon subtraction}
\author{Stefano Olivares\footnote{Stefano.Olivares@mi.infn.it} 
and Matteo G. A. Paris\footnote{Matteo.Paris@fisica.unimi.it}}
\affiliation{Dipartimento di Fisica, Universit\`a degli Studi di 
Milano, Italia}
\begin{abstract}
We address conditional de-Gaussification of continuous variable 
states by inconclusive photon subtraction (IPS) and review in 
details its application to bipartite twin-beam state of radiation. 
The IPS map in the Fock basis has been derived, as well as its 
counterpart in the phase-space. Teleportation assisted by IPS 
states is analyzed and the corresponding fidelity evaluated as 
a function of the involved parameters. Nonlocality of IPS states 
is investigated by means of different tests including displaced 
parity, homodyne detection, pseudospin, and displaced on/off 
photodetection. Dissipation and thermal noise are taken into account, 
as well as non unit quantum efficiency in the detection stage. 
We show that the IPS process, for a suitable choice of the 
involved parameters, improves teleportation fidelity and enhances 
nonlocal properties.
\end{abstract}
\maketitle
\section{Introduction} 
Nonclassical properties of the radiation field play a relevant
role in modern information processing since, in general, improve
continuous variable (CV) communication protocols based on light
manipulation \cite{vLB_rev,FOP:napoli:05}. Indeed, {\em quantum light}
finds application in several fundamental tests of quantum
mechanics \cite{ts0}, as well as in high precision measurements
and high capacity communication channels \cite{qcm, furusawa}.  
Among nonclassical features, entanglement plays a major role,
being the essential resource for quantum
computing, teleportation, and cryptographic protocols. Recently,
CV entanglement has been proved as a valuable tool also for
improving optical resolution, spectroscopy, interferometry,
tomography, and discrimination of quantum operations.
Recent experimental realizations also include dense coding
\cite{dense} and teleportation network \cite{qtn}.
\par
Entanglement in optical systems is usually generated through
parametric downconversion in nonlinear crystals. The resulting
bipartite state, the so-called twin-beam state of radiation
(TWB), allows the realization of several beautiful experiments 
and the demonstration of the above quantum protocols.  However, 
the resources available to generate CV entangled states are
unavoidably limited: nonlinearities are generally small, and, 
in turn, the resulting states have a limited amount of 
entanglement and energy. In this context, practical applications 
require novel schemes to create more entangled states or to 
increase the degree of entanglement of a given signal. 
\par
In quantum mechanics, the reduction postulate provides an
alternative mechanism to achieve {\em effective} nonlinear 
dynamics.  In fact, if a measurement is performed on a
portion of a composite system the output state strongly
depends on the results of the measurement. As a consequence, 
the {\em conditional} state of the unmeasured part, {\em i.e.}
the sub-ensemble corresponding to a given outcome, may be connected 
to the initial one by a (strongly) nonlinear map.
In this paper, we focus our attention on a scheme of this kind, 
and address a conditional method based on subtraction of photons
to enhance nonclassical features. In particular, we analyze 
how, and to which extent, photon subtraction may be used 
to increase nonlocal correlations of twin-beams.
As we will see, photon subtraction transforms the
Gaussian Wigner function of TWB into non-Gaussian one, and
therefore it is also referred to as a  {\em de-Gaussification} 
process.
\par
The photon subtraction process on TWBs was first proposed in
\cite{opatr}, where a well defined number of photons is being
subtracted from both the parties of a TWB, by transmitting each
mode through beam splitter and performing a joint photon-number
measurement on the reflected beams.  The degree of entanglement
is then increased and the the fidelity of the CV teleportation
assisted by such photon subtracted state is improved \cite{coch}.
However, this scheme is based on the possibility of resolving the
actual number of revealed photons. In \cite{ips:PRA:67} we showed
that the improvement of teleportation fidelity is possible also
when the number of detected photon is not known. In our scheme we
use on/off avalanche photodetectors able only to distinguish the
presence from the absence of radiation. For this reason we
referred to this method as to inconclusive photon subtraction
(IPS). The single-mode version of this process has been recently
implemented \cite{weng:PRL:04} and the nonclassicality of the
generated state starting from squeezed vacuum has been
theoretically investigated \cite{jeong,fock:oli}. 
In addition, nonlocal properties of the photon-subtracted TWBs have 
been investigated by means of
different nonlocality tests
\cite{ips:nonloc,OP:PSnoise,nha,sanchez,daffer,IOP:05}, finding
enhanced nonlocal properties depending on the particular test and
on the choice of the involved parameters.
\par
This paper is devoted to review the effects of IPS process on TWBs 
either in the ideal case, i.e., when the detection are not affected by
losses and no dissipation or thermal noise occurs during the
propagation of the involved modes, or when non unit quantum
efficiency is taken into account as well as the dynamics through
a noisy channel is considered. 
\par
The paper is structured as follows: in the next section 
we introduce photon subtraction as a method to enhance nonclassicality
of a radiation state and illustrate inconclusive photon subtraction
on a single-mode field. The de-Gaussification process
on two-mode fields is described in Sec.~\ref{s:IPS},
where the map of the IPS process is given both in the Fock representation
and in the phase-space.  In Sec.~\ref{s:dyn:TWB} we briefly review the dynamics
of a TWB in noisy channels and show that IPS can be profitably applied also
in the presence of noise. The CV
teleportation protocol is described in Sec.~\ref{s:tele}, where
we compare the teleportation fidelity when the protocol is assisted 
or not by the IPS process. In the following Sections, in order to 
characterize in details the nonlocal properties of the IPS states, 
we consider different {\em Bell} tests, namely, the nonlocality
test in the phase space (Sec.~\ref{s:DP}), the homodyne detection
test (Sec.~\ref{s:HD}), the pseudospin test (Sec.~\ref{s:PS}),
and a nonlocality test based on on/off photodetection
(Sec.~\ref{s:on:off}).  Finally, Sec.~\ref{s:remarks} closes 
the paper with some concluding remarks.
\section{Photon subtraction}\label{s:ps}
The idea of enhancing nonclassical properties of radiation by subtraction
of photons has been introduced in the context of Schr\"odinger cat
generation \cite{dak97} and subsequently applied to the improvement of CV
teleportation fidelity \cite{opatr}.  In the schemes of
Refs.~\cite{dak97,opatr} the field-mode to be ``photon subtracted'' (PS) is
impinged onto a beam-splitter with high transmissivity and whose second
port is left unexcited. At the output of the beam splitter the reflected
mode undergoes photon number measurement whereas the conditional state of
the transmitted mode represents the PS state. The properties of the PS
state depend on the number of detected photons, with single-photon
subtracted states that play a major role in the enhancement of
nonclassicality.  Unfortunately, the realization of photon number resolving
detectors is still experimentally challenging, and therefore a question
arises concerning the experimental feasibility of subtraction schemes. 
\par
Photodetectors that are usually available in quantum optics such as
avalanche photodiodes (APDs) operates in the Geiger mode
\cite{rev, serg}. They can be used to reconstruct the photon
statistics \cite{CVP,CVP:B} but cannot be used as photon counters. 
APDs show high quantum efficiency but their breakdown current 
is independent of the number of detected photons, which in turn 
cannot be determined. The outcome of these APD's is either ``off'' 
(no photons detected) or ``on'', i.e., a ``click'', indicating 
the detection of one or more photons. Actually, such an
outcome can be provided by any photodetector (photomultiplier,  
hybrid photodetector, cryogenic thermal detector) for which the 
charge contained in dark pulses is definitely below that of the output
current pulses corresponding to the detection of at least one
photon. Note that for most high-gain photmultipliers the anodic pulses
corresponding to no photons detected can be easily discriminated by
a threshold from those corresponding to the detection of one or more
photons.
\par
It appears therefore of interest to investigate the properties of 
photon subtracted states when the number of detected photons
is not discriminated. Such a process will be referred to as 
inconclusive photon subtraction (IPS) throughout the paper.
\begin{figure}
\begin{center}
\includegraphics[scale=0.98]{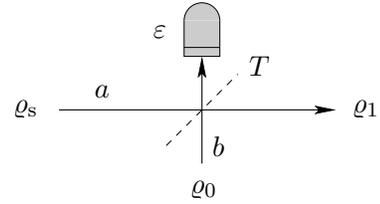}
\end{center}
\vspace{-.7cm}
\caption{Scheme of the IPS process: the input state
$\varrho_{\rm s}$ is mixed with the vacuum state $\varrho_{0} =
\ket{0}\bra{0}$ at a beam splitter (BS) with transmissivity $T$; then,
on/off photodetection with quantum efficiency $\varepsilon$ is performed
on the reflected beam. When the detector clicks we obtain the IPS state
$\varrho_1$\label{f:scheme}}
\end{figure}
The scheme of the IPS process is sketched in figure \ref{f:scheme}.  The 
mode $a$, excited in the state $\varrho_{\rm s}$ is mixed with the vacuum 
$\varrho_{0} = \ket{0}\bra{0}$ (mode $b$) at an unbalanced beam splitter (BS) 
with transmissivity $T=\cos^2\phi$ and then, on/off avalanche
photodetection with quantum efficiency $\varepsilon$ is performed on the
reflected beam.  APDs can only discriminate the presence of
radiation from the vacuum. The positive operator-valued measure
(POVM) $\{\Pi_0(\varepsilon),\Pi_1(\varepsilon)\}$ of the detector 
is given  by
\begin{eqnarray}
\Pi_0 (\varepsilon) = \sum_{k=0}^\infty(1-\varepsilon)^k \: |k\rangle\langle
k|\,, \qquad  \Pi_1 (\varepsilon) =  {\mathbb I} -  
\Pi_0(\varepsilon)\label{onoffPOM}\;.
\end{eqnarray}
The whole process can be characterized by $T$ and $\varepsilon$ which will 
be referred to as the IPS transmissivity and the IPS quantum efficiency.
The conditional state of the transmitted mode after the observation 
of a click is given by 
\begin{eqnarray}
\varrho_1 = \frac{1}{p_1(\phi,\varepsilon)} \hbox{Tr}_b
\left[U_{ab}(\phi)\varrho_{\rm s}\otimes\varrho_0\,U_{ab}^{\dag}(\phi)\: 
{\mathbb I}_a \otimes \Pi_1 (\varepsilon)\right]
\label{cond1}\;,
\end{eqnarray}
where $U_{ab}(\phi) = \exp\{-\phi(a^{\dag}b - a b^{\dag})\}$ is the
evolution operator of the beam splitter, and $p_1(\phi,\varepsilon)$ is the 
probability of a click. In general, the transformation (\ref{cond1})
realizes a non unitary quantum operation
$\varrho_1=\mathcal{E}(\varrho_{\rm s})$ with operator-sum decomposition
given by
\begin{eqnarray}
\mathcal{E}(\varrho_{\rm s})
= \frac{1}{p_1(\phi,\varepsilon)}\:
\sum_{p=1}^{\infty}\:m_p(\phi,\varepsilon)\:E_p(\phi)\: \varrho_{\rm s} \:
E_p^{\dag}(\phi)\:\label{map1}\;
\end{eqnarray}
where
\begin{align}
&m_p(\phi,\varepsilon) =
{\displaystyle \frac{\tan^{2p}\phi\,\,[1-(1-\varepsilon)^p]}{p!}}\,,
\\
&M_{p}(\phi) = a^p \, \cos^{a^\dag a}\phi \,.
\end{align}
which is found by explicit evaluation of the partial trace in 
(\ref{cond1}).
The IPS state obtained by applying the map (\ref{map1}) to a 
Gaussian state is no longer Gaussian, and therefore IPS represents
an effective source of non Gaussian states, which should be otherwise
generated by highly nonlinear, and thus inherently low rate, optical 
processes.
\par
In general the IPS process can produce an output state whose energy is
larger than the one of the input state and whose nonclassical properties
can be enhanced.  As an example, we address the photon subtraction onto a
Gaussian state described by the following Wigner function (using the Wigner
function formalism makes analytical calculations more straightforward):
\begin{equation}\label{single:gauss}
W_{\rm s}(z) = \frac{\exp\{ -F|z|^2 - G (z^2 + {z^*}^2) \}}
{\pi \sqrt{(F^2 - 4 G^2)^{-1}}}\,,
\end{equation}
whose energy is given by
\begin{equation}
E_{\rm s} = \int_{\mathbb{C}} d^2 z\, \left(|z|^2 - \frac12\right)\,
W_{\rm s}(z) = \frac{F}{F^2-4G^2} - \frac12\,.
\end{equation}
When the state (\ref{single:gauss}) undergoes the IPS process described
above, the Wigner function associated with the output state $\varrho_{1}$
reads \cite{fock:oli}
\begin{equation}\label{single:ips}
W_{1}(z) =
\frac{1}{p_1(\phi,\varepsilon)}\,\sum_{k=1}^{2}
\mathcal{C}_k(\phi,\varepsilon)\,
W_{k}(z)\,,
\end{equation}
with $\mathcal{C}_1(\phi,\varepsilon)=1$, $\mathcal{C}_2(\phi,\varepsilon)
= -(\varepsilon\sqrt{{\rm Det}[{\boldsymbol B} + \bmsigma_{\rm M}]})^{-1}$,
where
\begin{align}
{\boldsymbol B} = (1-T)\bmsigma + \frac{T}{2} \mathbbm{1}_2\,, \quad
\bmsigma_{\rm M} = \frac{2-\varepsilon}{2\varepsilon}\, \mathbbm{1}_2\,,
\end{align}
$\mathbbm{1}_2$ being the $2\times 2$ identity matrix, and $\bmsigma$ is
covariance associated with the state (\ref{single:gauss})
\begin{equation}
\bmsigma = \left(\begin{array}{cc}
(F+2G)^{-1} & 0 \\
0 & (F-2G)^{-1} 
\end{array}\right)\,,
\end{equation}
where $[\bmsigma]_{hk} = \frac12 \langle \{R_h,R_k\} \rangle -
\langle R_h \rangle \langle R_k \rangle$, $\{A,B\} = AB + BA$ denotes the
anticommutator, and
\begin{equation}
\boldsymbol{R} = (R_1,R_2)^T \equiv \left( \frac{a + a^{\dag}}{\sqrt{2}},
\frac{a - a^{\dag}}{i\sqrt{2}} \right)^T \,,
\end{equation}
$(\cdots)^{T}$ being the transposition operation.
Notice that $W_1(z)$
is no longer Gaussian. In Eq.~(\ref{single:ips}) we defined
\begin{equation}
W_{k}(z) = \frac{\exp\{ -F_k |z|^2 - G_k (z^2 + {z^*}^2) \}}
{\pi \sqrt{(F_k^2 - 4 G_k^2)^{-1}}}\,,
\end{equation}
where
\begin{align}
&F_1 = \mathcal{U}_{+} + \mathcal{U}_{-}\,, \quad
G_1 = \frac12(\mathcal{U}_{+} - \mathcal{U}_{-})\,, \\
&F_2 = 2(\mathcal{V}_{+} + \mathcal{V}_{-})\,, \quad
G_2 = \mathcal{V}_{+} - \mathcal{V}_{-}\,,
\end{align}
with
\begin{align}
&\mathcal{U}_{\pm} = \frac{F \pm 2G}
{2T+(1-T)(F \pm 2G)}\,, \\
&\mathcal{V}_{\pm} = \frac{F + 2(1 \pm G) T}
{4T + (1-T)[2\varepsilon + (2-\varepsilon)(F \pm 2 G)]\}}\,.
\end{align}
Because of the analytical expression (\ref{single:ips}), the energy of
the photon subtracted state is simply given by
\begin{eqnarray}
E_1(T,\varepsilon) = \frac{1}{p_1(T,\varepsilon)}\,\sum_{k=1}^{2}
\mathcal{C}_k\,\left[
\frac{F_k}{F_k^2-4G_k^2} - \frac12
\right]\,, 
\end{eqnarray}
with $\mathcal{C}_k \equiv \mathcal{C}_k(T,\varepsilon)$ and
where we put $T = \cos^2 \phi$.
\begin{figure}[tb]
\begin{center}
\includegraphics{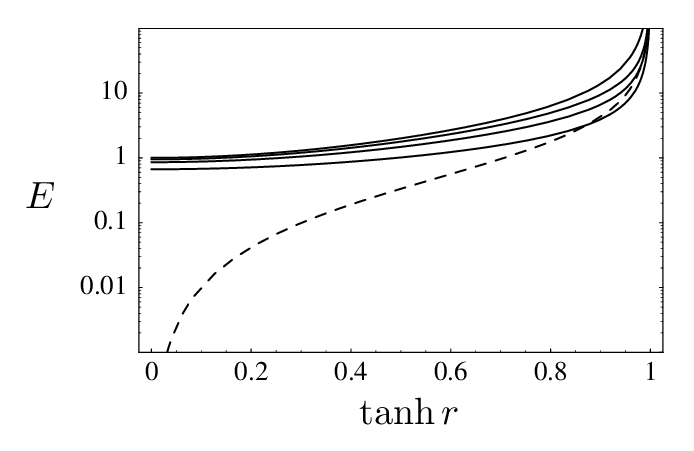}
\end{center}
\vspace{-0.5cm}
\caption{Logarithmic plots of the energies $E_{\rm s}$ (dashed line) and
$E_{1}$ (solid lines) in the case of a squeezed vacuum $\ket{0,r}$ as input
state and as functions of $\tanh r$ for $\varepsilon = 1$ and different
values of $T$.  From top to bottom (solid lines): $T=1$, $0.9$,  $0.75$,
and $0.5$.} \label{f:single:en}
\end{figure}
\begin{figure}[tb]
\begin{center}
\includegraphics{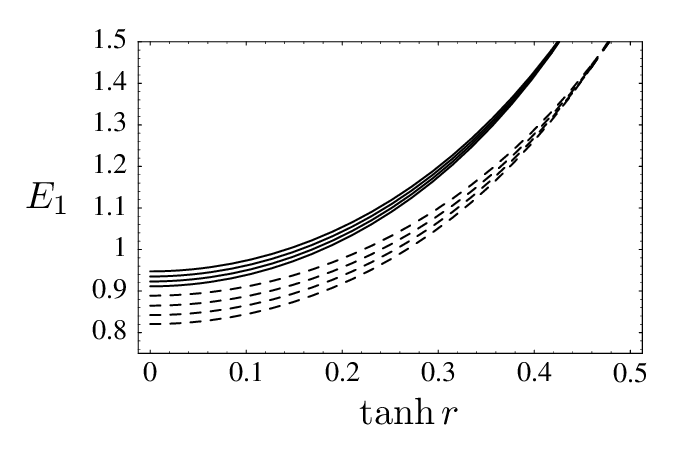}
\end{center}
\vspace{-0.5cm}
\caption{Plots of the energy $E_{1}$ of the IPS state in the case of a
squeezed vacuum $\ket{0,r}$ as input state as a function of $\tanh r$ for
$T = 0.9$ (solid lines) and $T = 0.8$ (dashed lines)
and different values of $\varepsilon$. From top to bottom: $\varepsilon=1$,
$0.75$,  $0.5$, and $0.25$.}
\label{f:single:eta}
\end{figure}
\par
Let us now focus our attention on the IPS process applied to the squeezed
vacuum $\ket{0,r} = S(r)\ket{0}$, $S(r) =
\exp\{ \frac12 r ({a^{\dag}}^2 - a^2) \}$ being the squeezing operator,
which has been recently realized experimentally \cite{weng:PRL:04}.
The Wigner function associated with $\ket{0,r}$ is given by
Eq.~(\ref{single:gauss}) with $F = 2 \cosh 2r$ and $G = - \sinh 2r$.
In Figs.~\ref{f:single:en} we plot the energies
$E_{\rm s}$ and $E_{1}$ of the input and output states, respectively, for
different values of the involved parameters as functions of $\tanh r$.
We can see that there is a threshold on $r$, depending on $T$ and
$\varepsilon$, under which the IPS state has a larger energy than the input
state.  Furthermore, when $\varepsilon = 1$, $T \to 1$ and $r \to 0$ we can
see that $E_1 \to 1$: in these limits the output state approaches to the
squeezed Fock state $S(r)\ket{1}$ \cite{jeong,fock:oli}.
Finally, in Fig.~\ref{f:single:eta} $E_1$ is plotted for two values of
$T$ and different values of $\varepsilon$ as a function of $\tanh r$: we
find that as $r$ increases, the IPS efficiency is not so relevant in the
process.
\section{Photon subtraction on bipartite states}\label{s:IPS}
\begin{figure}[tb]
\begin{center}
\includegraphics[scale=0.98]{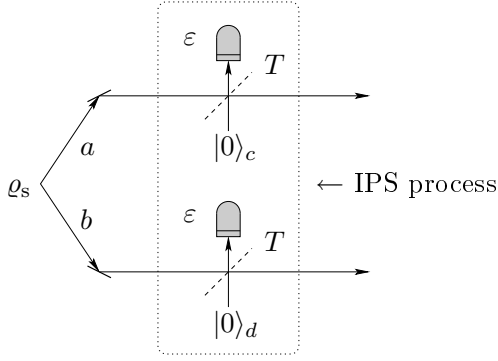}
\end{center}
\vspace{-0.5cm}
\caption{Scheme of the IPS process. The two modes, $a$ and $b$, of a shared
bipartite state $\varrho_{\rm s}$ are mixed with the vacuum at two BSs
with equal transmissivity $T$ and on/off photodetection with quantum
efficiency $\varepsilon$ is performed on the reflected beam: when both the
detectors click one obtains the IPS state.}\label{f:IPS:scheme}
\end{figure}
In this Section we address de-Gaussification of bipartite states by IPS . 
The de-Gaussification can be achieved by subtracting
photons from both modes through on/off detection \cite{ips:PRA:67,coch}. 
The IPS scheme for two modes is sketched in Fig.~\ref{f:IPS:scheme}. The
modes $a$ and $b$ of the shared bipartite state $\varrho_{\rm s}$ are mixed
with vacuum modes at two unbalanced beam splitters (BS) with equal
transmissivity $T = \cos^2\phi$; the  reflected modes $c$ and $d$ are then
revealed by avalanche photodetectors (APD) with equal efficiency
$\varepsilon$.  The conditional measurement on modes $c$ and $d$, is
described by the POVM (assuming equal quantum efficiency for the
photodetectors)
\begin{align}
{\Pi}_{00} (\varepsilon) &= {\Pi}_{0,c} (\varepsilon) \otimes {\Pi}_{0,d}
(\varepsilon)\,, \\
{\Pi}_{01} (\varepsilon) &= {\Pi}_{0,c} (\varepsilon) \otimes
{\Pi}_{1,d} (\varepsilon)\:, \\
{\Pi}_{10} (\varepsilon) &= {\Pi}_{1,c} (\varepsilon) \otimes {\Pi}_{0,d}
(\varepsilon)\,, \\
{\Pi}_{11} (\varepsilon) &= {\Pi}_{1,c} (\varepsilon) \otimes
{\Pi}_{1,d} (\varepsilon)\;.
\label{povm11}
\end{align}
When the two photodetectors jointly click, the conditioned output state
of modes $a$ and $b$ is given by \cite{ips:PRA:67,ips:nonloc}
\begin{widetext}
\begin{equation}
\mathcal{E}(\varrho_{\rm s})
= \frac{\hbox{Tr}_{cd}\big[
U_{ac}(\phi)\otimes U_{bd}(\phi) \: \varrho_{\rm s}
\otimes |0\rangle_{cd}{}_{dc}\langle 0|
\: U_{ac}^{\dag}(\phi)\otimes U_{bd}^{\dag}(\phi) \:
{\mathbb I}_a \otimes
{\mathbb I}_b \otimes
{\Pi}_{11} (\varepsilon)
\big]}{p_{11}(r,\phi,\varepsilon)}\:, \label{ptr}
\end{equation}
\end{widetext}
where $U_{ac}(\phi)=\exp\{-\phi(a^{\dag} c-a c^{\dag}) \}$ and
$U_{bd}(\phi)$ are the evolution operators of the beam splitters,
$|0\rangle_{cd} \equiv |0\rangle_{c}\otimes|0\rangle_{d}$, and
$p_{11}(r,\phi,\varepsilon)$ is the probability of a click
in both the detectors.
The partial trace on modes $c$ and $d$ can be explicitly evaluated, thus
arriving at the following decomposition of the IPS map:
\begin{multline}
\mathcal{E}(\varrho_{\rm s})
= \frac{1}{p_{11}(r,\phi,\varepsilon)}\:\\
\times \sum_{p,q=1}^{\infty}\:m_p(\phi,\varepsilon)
\:M_{pq}(\phi)\: \varrho_{\rm s}
\: M_{pq}^{\dag}(\phi)\: m_q(\phi,\varepsilon)\:\label{KE}
\end{multline}
where
\begin{equation}
M_{pq}(\phi) = \frac{\mbox{}}{\mbox{}}
a^p b^q \, (\cos\phi)^{a^\dag a + b^\dag b}\,.
\end{equation}
Eq.~(\ref{KE}) is indeed an operator-sum representation of the IPS map:
$\{p,q\}\equiv \theta$ should be intended as a polyindex so that (\ref{KE})
reads $\mathcal{E}(\varrho_{\rm s}) = \sum_\theta A_\theta \varrho_{\rm s}
A^\dag_\theta$ with $A_\theta= [p_{11}(r,\phi,\varepsilon)]^{-1/2}\,
m_p(\phi,\varepsilon)\:M_{pq}(\phi)$.
\par
From now on we focus our attention on the case in which the shared state is
the twin-beam state of radiation (TWB) $\varrho_{\rm s} =
\dket{\Lambda}\dbra{\Lambda}$, where $\dket{\Lambda} =
\sqrt{1-\lambda^2}\sum_k \lambda^{k} \ket{k}\otimes\ket{k}$ with
$\lambda=\tanh r$, $r$ being the TWB squeezing parameter. The TWB is
obtained by parametric down-conversion of the vacuum, $\dket{\Lambda} =
\exp\{ r(a^\dag b^\dag - ab) \}\ket{0}$, $a$ and $b$ being field operators,
and it is described by the Gaussian Wigner function
\begin{equation}
W_{0}(\alpha,\beta) =
\frac{\exp\{
-2 \widetilde{A}_0 (|\alpha|^2+|\beta|^2)
+ 2 \widetilde{B}_0 (\alpha\beta + \calpha\cbeta) \}}
{\pi^2\sqrt{{\rm Det}[\bmsigma_0]}}\,,
\label{twb:wig}
\end{equation}
with
\begin{equation}
\widetilde{A}_0 = \frac{A_0}{4 \sqrt{{\rm Det}[\bmsigma_0]}}\,,\qquad
\widetilde{B}_0 = \frac{B_0}{4 \sqrt{{\rm Det}[\bmsigma_0]}}\,,
\end{equation}
where $A_0 \equiv A_0(r) = \cosh(2 r)$,
$B_0 \equiv B_0(r) = \sinh (2 r)$ and $\bmsigma_0$ is the covariance matrix
\begin{equation}\label{cvm:twb}
\bmsigma_0 = \frac12
\left(
\begin{array}{cc}
A_0\, \mathbbm{1}_2 & B_0\, \bmsigma_3\\[1ex]
B_0\, \bmsigma_3 & A_0\, \mathbbm{1}_2
\end{array}\right)\:,
\end{equation}
$\mathbbm{1}_2$ being the $2 \times 2$ identity matrix and $\bmsigma_3 =
{\rm Diag}(1,-1)$; $\bmsigma_0$ is defined as $[\bmsigma_0]_{hk} =
\frac12\langle \{ R_h, R_k \}\rangle - \langle R_h \rangle
\langle R_k \rangle$ with
\begin{align}
\boldsymbol{R} &= (R_1, R_2, R_3, R_4)^{T} \\
&\equiv \left(
\frac{a+a^{\dag}}{\sqrt{2}},\frac{a-a^{\dag}}{i\sqrt{2}},
\frac{b+b^{\dag}}{\sqrt{2}},\frac{b-b^{\dag}}{i\sqrt{2}}
\right)^{T}\,.
\end{align}
Now we explicitly calculate the Wigner function of
the corresponding IPS state, which, as one may expect, is no longer
Gaussian and positive-definite. The state entering the two beam splitters
is described by the Wigner function
\begin{equation}
W_{0}^{\hbox{\tiny (in)}}(\alpha,\beta,\zeta,\xi) =
W_{0}(\alpha,\beta)\,
\frac{4}{\pi^2} \exp\left\{ -2|\zeta|^2 - 2|\xi|^2 \right\}\,,
\end{equation}
where the second factor at the right hand side represents the two vacuum
states of modes $c$ and $d$.
The action of the beam splitters on $W^{\hbox{\tiny (in)}}_{r}$ can be
summarized by the following change of variables \cite{FOP:napoli:05}
\begin{align}
&\alpha \to \alpha\cos\phi + \zeta\sin\phi\,,
&\zeta  \to \zeta\cos\phi  - \alpha\sin\phi\,,\\
&\beta  \to \beta\cos\phi  + \xi\sin\phi\,,
&\xi    \to \xi\cos\phi    - \beta\sin\phi\,,
\end{align}
and the output state, after the beam splitters, is then given by
\begin{multline}
W_{r,\phi}^{\hbox{\tiny (out)}}(\alpha,\beta,\zeta,\xi) =\\
\frac{4}{\pi^2}\, W_{r,\phi}(\alpha,\beta)\,
\exp\left\{ -a |\xi|^2 + w \xi + \cw \cxi \right\} \\
\times\exp\big\{ -a |\zeta|^2  + (v + 2 \witB_0 \xi \sin^2\phi)\zeta \\
+ (\cv + 2 \witB_0 \cxi \sin^2\phi)\czeta \big\}\,,
\end{multline}
where
\begin{multline}
W_{r,\phi}(\alpha,\beta) =\\
\frac{
\exp\{ -b (|\alpha|^2 + |\beta|^2)
+ 2 \witB_0 \cos^2\phi\, (\alpha\beta + \calpha\cbeta) \}}
{\pi^2\sqrt{{\rm Det}[\bmsigma_0]}}
\end{multline}
and
\begin{align}
&a \equiv a(r,\phi) = 2 (\witA_0 \sin^2\phi + \cos^2\phi),\\
&b \equiv b(r,\phi) = 2 (\witA_0 \cos^2\phi + \sin^2\phi)\,,\\
&v \equiv v(r,\phi) = 2 \cos\phi\, \sin\phi\,
[(1-\witA_0)\calpha + \witB_0 \beta],\\
&w \equiv w(r,\phi) = 2 \cos\phi\, \sin\phi\,
[(1-\witA_0)\cbeta + \witB_0 \alpha]\,.
\end{align}
\par
At this stage on/off detection is performed on modes
$c$ and $d$ (see Fig.~\ref{f:IPS:scheme}). We are interested in
the situation when both the detectors click. The Wigner function
of the double click element $\Pi_{11}(\varepsilon)$ of the POVM
[see Eq.~(\ref{povm11})] is given by \cite{ips:PRA:67,cond:cola}
\begin{align}
W_{\varepsilon}(\zeta,\xi) &\equiv
W[\Pi_{11}(\varepsilon)](\zeta,\xi)\\
&= \frac{1}{\pi^2}\{
1-Q_{\varepsilon}(\zeta)-Q_{\varepsilon}(\xi)
+Q_{\varepsilon}(\zeta) Q_{\varepsilon}(\xi)
\}\,,
\end{align}
with
\begin{equation}
Q_{\varepsilon}(z) = \frac{2}{2-\varepsilon}\,
\exp\Bigg\{-\frac{2\varepsilon}{2-\varepsilon}\, |z|^2 \Bigg\}\,.
\end{equation}
Using Eq.~(\ref{ptr}) and the phase-space expression of trace
for each mode, i.e.,
\begin{equation}
{\rm Tr}[O_1\,O_2] = \pi \int_{\mathbb C} d^2z\, W[O_{1}](z)\,
W[O_{2}](z)\,,
\end{equation}
the Wigner function of the output state, conditioned to the double click
event, reads
\begin{equation}\label{w:ips:informal}
W_{r,\phi,\varepsilon}(\alpha,\beta) =
\frac{f(\alpha,\beta)}{p_{11}
(r,\phi,\varepsilon)}\,,
\end{equation}
where $f(\alpha,\beta) \equiv f_{r,\phi,\varepsilon}(\alpha,\beta)$ with
\begin{multline}\label{w:ips:informal:f}
f(\alpha,\beta) =
\pi^2\,\int_{\mathbb{C}^2}d^2\zeta\,d^2\xi\,
\frac{4}{\pi^2}\,W_{r,\phi}(\alpha,\beta)\, \\
\times \sum_{k=1}^4 \frac{C_k}{\pi^2}\,
G_{r,\phi,\varepsilon}^{(k)}(\alpha,\beta,\zeta,\xi)\,,
\end{multline}
with $C_k \equiv C_k(\varepsilon)$ and $C_1 = 1$,
$C_2 = C_3 = -2(2-\varepsilon)^{-1}$, $C_4 = 4(2-\varepsilon)^{-2}$;
the double-click probability $p_{11}(r,\phi,\varepsilon)$
can be written as function of $f(\alpha,\beta)$ as follows
\begin{equation}\label{w:ips:informal:p}
p_{11}(r,\phi,\varepsilon) =
\pi^2\,\int_{\mathbb{C}^2}d^2\alpha\,d^2\beta\,
f(\alpha,\beta)\,.
\end{equation}
The quantities $G_{r,\phi,\varepsilon}^{(k)}(\alpha,\beta,\zeta,\xi)$ 
in Eq.~(\ref{w:ips:informal:f}) are given by 
\begin{multline}
G_{r,\phi,\varepsilon}^{(k)}(\alpha,\beta,\zeta,\xi) = \\
\exp\big\{ -x_k |\zeta|^2 + (v + 2 B \xi \sin^2\phi)\zeta \\
+(\cv + 2 B \cxi \sin^2\phi)\czeta \big\} \\
\times\exp\left\{ -y_k |\xi|^2 + w \xi + \cw \cxi \right\}\,,
\label{meas:int}
\end{multline}
where $x_k \equiv x_k(r,\phi,\varepsilon)$, and
$y_k \equiv y_k(r,\phi,\varepsilon)$ are
\begin{align}
&x_1 = x_3 = y_1 = y_2 = a \nonumber\\
&x_2 = x_4 = y_3 = y_4 = a + 2\varepsilon(2-\varepsilon)^{-1}\,. \nonumber
\end{align}
After the integrations we have
\begin{multline}
f(\alpha,\beta) =\frac{1}{\pi^2}\,
\sum_{k=1}^4 {\cal C}_k
\,\exp\{ (f_k-b)|\alpha|^2 + (g_k-b)|\beta|^2 \\
+(2 \witB_0 T + h_k)(\alpha\beta + \calpha\cbeta)\}
\end{multline}
and
\begin{equation}
p_{11}(r,T,\varepsilon) =
\sum_{k=1}^4
\frac{{\cal C}_k}
{(b-f_k)(b-g_k)-(2 \witB_0 T + h_k)^2}
\,,
\end{equation}
where we put $T = \cos^2\phi = 1 - \sin^2 \phi$, and defined
\begin{equation}
{\cal C}_k \equiv {\cal C}_k (r,T,\varepsilon) =
\frac{4 C_k}
{[x_ky_k - 4 \witB_0^2 (1-T)^2]\sqrt{{\rm Det}[\bmsigma_0]}}
\end{equation}
and $f_k \equiv f_k(r,T)$, $g_k \equiv g_k(r,T)$, and $h_k \equiv h_k(r,T)$
given by
\begin{align}
&f_k =
{\cal N}_k
\, [x_k \witB_0^2 + 4 \witB_0^2 (1-\witA_0)(1-T) + y_k (1-\witA_0)^2]\,,\\
&g_k =
{\cal N}_k
\, [x_k (1-\witA_0)^2 + 4 \witB_0^2 (1-\witA_0)
(1-T) + y_k \witB_0^2]\,,\\
&h_k =
{\cal N}_k
\, \{(x_k + y_k) \witB_0 (1-\witA_0)\nonumber\\
&\hspace{2cm}
+ 2 \witB_0 [\witB_0^2 + (1-\witA_0)^2] (1-T)\}\,,\\
&{\cal N}_k \equiv  {\cal N}_k(r,T) =
{\displaystyle
\frac{4 T\, (1-T)}{x_k y_k - 4 \witB_0^2(1-T)^2}\,.
}
\end{align}
In this way, the Wigner function of the IPS state can be rewritten as
\begin{eqnarray}\label{ips:wigner}
W_{\hbox{\rm\tiny IPS}}(\alpha,\beta) =
\frac{1}{\pi^2\, p_{11}(r,T,\varepsilon)}
\sum_{k=1}^4 {\cal C}_k\,W_{k}(\alpha,\beta)\,,
\end{eqnarray}
with
\begin{multline}
W_{k}(\alpha,\beta) =
\exp\{ (f_k-b) |\alpha|^2 +(g_k-b) |\beta|^2 \\
+ (2\witB_0 T + h_k) (\alpha\beta + \calpha\cbeta)\}\,.
\end{multline}
Finally, the density matrix corresponding to $W_{\hbox{\rm\tiny
IPS}}(\alpha,\beta)$ reads as follows \cite{ips:PRA:67}
\begin{multline}
\label{ips:fck}
{\varrho}_{\hbox{\rm\tiny IPS}} =
\frac{1-\lambda^2}{p_{11}(r,T,\varepsilon)}
  \sum_{n,m=0}^{\infty} (\lambda T)^{n+m} 
  \sum_{h,k=0}^{{\rm Min}[n,m]}  {\cal A}_{h,k}(T,\varepsilon) \\
  \times
  \sqrt{ \frac{n}{h}\frac{n}{k}\frac{m}{h}\frac{m}{k} }
  \, \ket{n-k}_a \ket{n-h}_b {_b}\bra{m-h} {_a}\bra{m-k}\,,
\end{multline}
where $\lambda = \tanh r$ and
\begin{equation}\label{ips:fhk}
  {\cal A}_{h,k}(T,\varepsilon)= \left[ 1 - (1-\varepsilon)^h \right]
  \left[ 1 - (1-\varepsilon)^k
  \right] \left( \frac{1-T}{T} \right)^{h+k}\:.
\end{equation}
\par
\begin{figure}[tb]
\begin{center}
\includegraphics{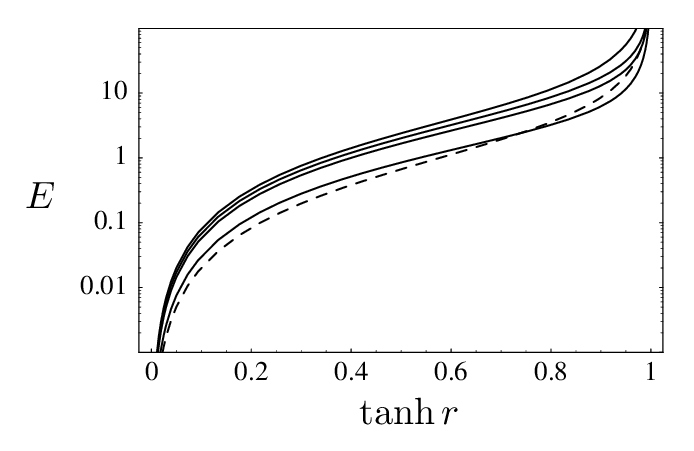}
\end{center}
\vspace{-0.5cm}
\caption{Logarithmic plots of the energies $E_{\rm s}$ (dashed line) and
$E_{\hbox{\rm\tiny IPS}}$ (solid lines) in the case of a TWB as input state
as functions of $\tanh r$ for $\varepsilon = 1$ and different values of
$T$.  From top to bottom (solid lines): $T=1$, $0.9$,  $0.75$, and $0.5$.}
\label{f:ips:en}
\end{figure}
\begin{figure}[tb]
\begin{center}
\includegraphics{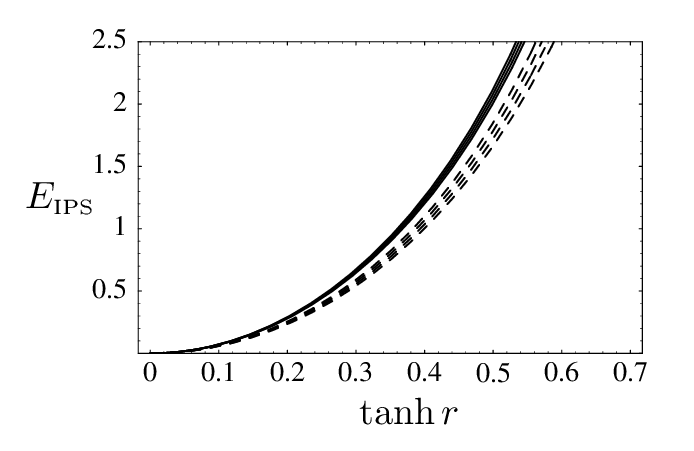}
\end{center}
\vspace{-0.5cm}
\caption{Plots of the energy $E_{\hbox{\rm\tiny IPS}}$ of the IPS state in
the case of the TWB as input state as a function of $\tanh r$ for
$T = 0.9$ (solid lines) and $T = 0.8$ (dashed lines)
and different values of $\varepsilon$. From top to bottom: $\varepsilon=1$,
$0.75$,  $0.5$, and $0.25$.}
\label{f:ips:eta}
\end{figure}
In Fig.~\ref{f:ips:en} we plot the energies $E_{\rm s}$ and
$E_{\hbox{\rm\tiny IPS}}$ of the bipartite input and output states,
respectively, for different values of the involved parameters as functions
of $\tanh r$. We recall that for a given Wigner function $W(v,w)$ of a
bipartite state, the corresponding energy is
\begin{equation}
E = \int_{\mathbb{C}^2} d^2 v\, d^2 w\, \left(|v|^2 + |w|^2 - 1\right)\,
W(v,w)\,.
\end{equation}
If the bipartite state has a Wigner function of the form
\begin{equation}\label{two:generic}
W_{\rm s}(v,w) =
\frac{\exp\{ -F|v|^2 -G|w|^2 + H (vw + v^* w^*) \}}
{\pi^2 (FG-H^2)^{-1}}\,,
\end{equation}
then its energy reads:
\begin{equation}
E_{\rm s} = \frac{F+G}{2(FG - H^2)}-1\,;
\end{equation}
thereby, in the case of a TWB as input state $F$, $G$, and $H$ are
obtained from Eq.~(\ref{twb:wig}) and the energy of the state emerging
from the IPS process can be written as
\begin{equation}
E_{\hbox{\rm\tiny IPS}} = \frac{1}{\pi^2\, p_{11}(r,T,\varepsilon)}
\sum_{k=1}^4 {\cal C}_k \left[
\frac{F_k+G_k}{2(F_k G_k - H_k^2)^2}-1
\right]\,
\end{equation}
with $F_k = b-f_h$, $G_k = b-g_h$, and $H_k = 2 \witB_0 T + h_k$ and all
the involved quantities are the same as in Eq.~(\ref{ips:wigner}).
As in the single mode case, we can see that there is a threshold on $r$,
depending on $T$ and $\varepsilon$, under which the IPS state has a larger
energy than the input state. In Fig.~\ref{f:ips:eta} $E_{\hbox{\rm \tiny
IPS}}$ is plotted for two values of $T$ and different values of
$\varepsilon$ as a function of $\tanh r$: we find that as $r$ decreases,
the IPS efficiency is not so relevant.
\par
The state given in Eq.~(\ref{ips:wigner}) is no longer a Gaussian state and
its use in the improvement of continuous variable teleportation
\cite{ips:PRA:67} as well as in the enhancement of the nonlocality
\cite{ips:nonloc,nha,sanchez} will be investigated in the following Sections.
\section{Dynamics of TWB in noisy channels}\label{s:dyn:TWB}
Before addressing the properties of the IPS bipartite state described in
the previous Section, we review the evolution of the twin-beam state of
radiation (TWB) in a noisy environment, namely, an environment where
dissipation and thermal noise take place \cite{OP:PSnoise}. As we
will see, we can include in our analysis the effect due to the propagation
through this kind of channel by a simple change of the involved quantities.
Using a more compact form, Eq.~(\ref{twb:wig}) can
also be rewritten as
\begin{equation}\label{gauss:form}
W_{0}(\bmX) =
\frac{\exp\left\{ -\frac12\, \bmX^{T}\,\bmsigma_{0}^{-1}\,\bmX \right\}}
{\pi^2 \sqrt{{\rm Det}[\bmsigma_0]}}\,,
\end{equation}
with $\bmX = (x_1,y_1,x_2,y_2)^{T}$, $\alpha=\frac{1}{\sqrt{2}}(x_1+iy_1)$
and $\beta=\frac{1}{\sqrt{2}}(x_2+iy_2)$, and $(\cdots)^{T}$ denoting the
transposition operation.
\par
When the two modes of the TWB interact with a noisy environment, namely in the
presence of dissipation and thermal noise, the evolution of the Wigner
function (\ref{twb:wig}) is described by the following Fokker-Planck equation
\cite{wm:quantopt:94,binary,seraf:PRA:69}
\begin{equation}\label{fp:eq:cmp}
\partial_t W_{t}(\bmX) = \frac12 \Big(
\partial_{\bmX}^T \bbGamma \bmX + \partial_{\bmX}^T
\bbGamma \bmsigma_{\infty} \partial_{\bmX} \Big) W_{t}(\bmX)\,,
\end{equation}
with $\partial_{\bmX} =
(\partial_{x_1},\partial_{y_1},\partial_{x_2},\partial_{y_2})^{T}$.
The damping matrix is given by $\bbGamma = \bigoplus_{k=1}^2\,
\Gamma_k \mathbbm{1}_2$, whereas
\begin{eqnarray}
\bmsigma_{\infty} &= \bigoplus_{k=1}^{2}\, \bmsigma_{\infty}^{(k)} =
\left(
\begin{array}{cc}
\bmsigma_{\infty}^{(1)} & \boldsymbol{0} \\[1ex]
\boldsymbol{0} & \bmsigma_{\infty}^{(2)}
\end{array}
\right)\,,
\end{eqnarray}
where $\boldsymbol{0}$ is the $2 \times 2$ null matrix and
\begin{equation}
\bmsigma_{\infty}^{(k)} = 
\frac12
\left(
\begin{array}{cc}
1 + 2 N_{k} & 0\\[1ex]
0 & 1 + 2 N_k
\end{array}
\right)\,.
\end{equation}
$\Gamma_k$, $N_k$ denote the damping rate and the average number of
thermal photons of the channel $k$, respectively. $\bmsigma_{\infty}$
represents the covariance matrix of the environment and, in turn, the
asymptotic covariance matrix of the evolved TWB.  Since the environment is
itself excited in a Gaussian state, the evolution induced by
(\ref{fp:eq:cmp}) preserves the Gaussian form (\ref{gauss:form}).  The
covariance matrix at time $t$ reads as follows
\cite{seraf:PRA:69,FOP:napoli:05}
\begin{equation}
\bmsigma_t = \mathbbm{G}_t^{1/2}\,\bmsigma_0\,\mathbbm{G}_t^{1/2}
+ (\mathbbm{1} - \mathbbm{G}_t)\,\bmsigma_{\infty}\,,
\end{equation}
where $\mathbbm{G}_t = \bigoplus_{k=1}^2\,e^{-\Gamma_k t}\,\mathbbm{1}_2$.
The covariance matrix $\bmsigma_t$ can be also written as
\begin{equation}\label{evol:cvm:12}
\bmsigma_t = \frac 12
\left(
\begin{array}{cc}
A_t(\Gamma_1,N_1)\, \mathbbm{1}_2&  B_t(\Gamma_1)\,\bmsigma_3 \\[1ex]
B_t(\Gamma_2)\, \bmsigma_3 & A_t(\Gamma_2,N_2)\, \mathbbm{1}_2 
\end{array}
\right)
\end{equation}
with
\begin{equation}
\label{AtBt}
\begin{array}{l}
A_t(\Gamma_k,N_k) = A_0\,e^{-\Gamma_k t}
+ \left(1-e^{-\Gamma_k t}\right) (1 + 2 N_k)\,,\\[1ex] 
B_t(\Gamma_k) = B_0\,e^{-\Gamma_k t}\,.
\end{array}
\end{equation}
Finally, if we assume $\Gamma_1 = \Gamma_2 = \Gamma$ and $N_1 = N_2 = N$,
then the covariance matrix (\ref{evol:cvm:12}) becomes formally identical
to (\ref{cvm:twb}) and the corresponding Wigner function reads
\begin{equation}
W_{t}(\alpha,\beta) =
\frac{\exp\{
-2 \widetilde{A}_t (|\alpha|^2+|\beta|^2)
+ 2 \widetilde{B}_t (\alpha\beta + \calpha\cbeta)\}}
{\pi^2\sqrt{{\rm Det}[\bmsigma_t]}}\,,
\label{twb:wig:noise}
\end{equation}
with
\begin{equation}
\widetilde{A}_t = \frac{A_t(\Gamma,N)}{4\sqrt{{\rm Det}[\bmsigma_t]}}\,,
\qquad  
\widetilde{B}_t = \frac{B_t(\Gamma)}{4\sqrt{{\rm Det}[\bmsigma_t]}}\,.
\end{equation}
If the IPS process is performed on a TWB evolved in a noisy
environment with both the channels having the same damping rate and thermal
noise, then the Wigner function of the state arriving at the beam splitters
is now given by Eq.~(\ref{twb:wig:noise}), and the output state is still
described by Eq.~(\ref{ips:wigner}), but with the following substitutions
\begin{equation}\label{sostituzioni}
\witA_0 \to \widetilde{A}_t \,,\quad
\witB_0 \to \widetilde{B}_t \,,\quad
\bmsigma_0 \to \bmsigma_t\,.
\end{equation}
\section{Continuous variable teleportation}\label{s:tele}
\begin{figure}[tb]
\begin{center}
\includegraphics[scale=0.98]{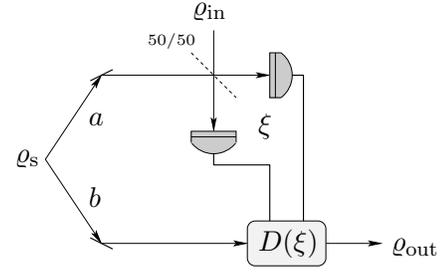}
\end{center}
\vspace{-0.5cm}
\caption{Scheme of the CV teleportation. One of the two modes of a shared
bipartite state $\varrho_{\rm s}$ is mixed with the input state
$\varrho_{\rm in}$ at a balanced BS, and  then a double homodyne detection
is performed on the two output modes measuring the complex outcome $\xi$.
The teleported state $\varrho_{\rm out}$ is obtained displacing by the same
amount $\xi$ the remaining mode of the shared state and averaging over all
the possible outcomes.}\label{f:TScheme}
\end{figure}
The scheme of continuous variable (CV) teleportation is sketched in
Fig.~\ref{f:TScheme}. A bipartite state $W_{\rm s}$ is shared between two
parties: one mode of the state is mixed at a balanced beam splitter (BS)
with the state to be teleported, $W_{\rm in}$, then double-homodyne
measurement is performed on the two emerging modes. The complex outcome
$\xi$ of the measurement is used in order to displace the remaining mode of
$W_{\rm s}$ and the teleported state $W_{\rm out}$ is obtained averaging
over all the possible outcomes.  Here we address the teleportation of the
coherent state $\ket{\alpha}$, whose Wigner function reads
\begin{equation}
W_{\rm in}(z) = \frac{2}{\pi} \exp\{ -2 |z-\alpha|^2 \}\,.
\end{equation}
If we consider the following generic shared state:
\begin{equation}
W_{\rm s}(v,w) =
\frac{\exp\{ -F|v|^2 -G|w|^2 + H (vw + v^* w^*) \}}
{\pi^2 (FG-H^2)^{-1}}\,,
\end{equation}
and since the POVM describing the double homodyne detection is
\begin{equation}
W_{\xi}(z,v) = \frac{1}{\pi^2}\, \delta^{(2)}(z-v^*-\xi)\,,
\end{equation}
$\delta^{(2)}(\zeta)$ being the complex Dirac's delta function, the output
state $W_{\rm out}$ is given by \cite{FOP:napoli:05}
\begin{align}
W_{{\rm out}}(w) &= \pi^2 \int_{\mathbb C}d^2\xi
\int_{{\mathbb C}^2}d^2z\,d^2v\,  W_{\rm in}(z)\,\nonumber \\
&\hspace{2cm} \times\,W_{\rm s}(v,w-\xi)\,
W_{\xi}(z,v)\\
&=\frac{1}{\pi \sigma_{\rm out}}
\exp\left\{ - \frac{|w-\alpha|^{2}}{\sigma_{\rm out}}\right\}\,,
\end{align}
where
\begin{equation}
\sigma_{\rm out} = \frac{1}{2} + \frac{F+G+2H}{FG-H^2}\,;
\end{equation}
in turn, the average fidelity of teleportation of coherent states reads as
follows:
\begin{align}\label{gen:fid}
\overline{F} &\equiv
\pi \int_{\mathbb C} d^2w\, W_{\rm in}(w)\,W_{\rm out}(w)\\
&=\frac{FG-H^2}{FG-H^2+F+G-2H} = \frac{2}{1+2\,\sigma_{\rm out}}\,.
\end{align}
\par
\begin{figure}[tb]
\begin{center}
\includegraphics{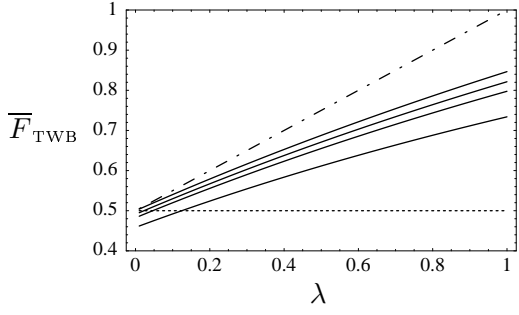}
\end{center}
\vspace{-0.5cm}
\caption{Plots of the teleportation fidelity $\overline{F}_{\mbox{\tiny
TWB}}$ assisted by TWB in the ideal case ($\Gamma t = N = 0$, dot-dashed
line) as a function of $\lambda = \tanh r$. The solid lines are
$\overline{F}_{\mbox{\tiny TWB}}$ with $\Gamma t = 0.2$ and, from top to
bottom, $N = 0$, $0.1$, $0.2$, and $0.5$.}\label{f:fid:TWB}
\end{figure}
\begin{figure}[tb]
\begin{center}
\includegraphics{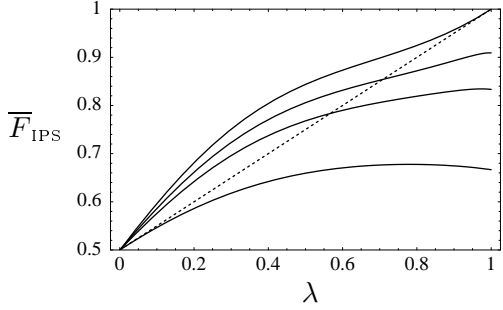}
\end{center}
\vspace{-0.5cm}
\caption{Plots of the teleportation fidelity $\overline{F}_{\mbox{\tiny
IPS}}$ assisted by IPS in the ideal case ($\Gamma t = N = 0$) as a function
of $\lambda = \tanh r$. The dashed line is $\overline{F}_{\mbox{\tiny
TWB}}$, whereas the solid lines are $\overline{F}_{\mbox{\tiny IPS}}$ with
$\varepsilon = 1$ and, from top to bottom, $T = 1.0$, $0.9$, $0.8$, and
$0.5$.}\label{f:fidid}
\end{figure}
\begin{figure}[tb]
\begin{center}
\includegraphics{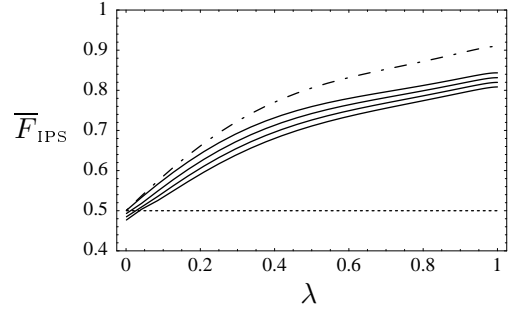}
\end{center}
\vspace{-0.5cm}
\caption{Plots of the teleportation fidelity $\overline{F}_{\mbox{\tiny
IPS}}$ assisted by IPS as a function of $\lambda = \tanh r$ with $T = 0.9$,
$\varepsilon = 1$, $\Gamma t = 0.1$ and different values of $N$ (solid
lines): from top to bottom $N = 0$, $0.1$, $0.2$, and $0.3$. The dot-dashed
line is $\overline{F}_{\mbox{\tiny IPS}}$ with $T = 0.9$, $\varepsilon =
1$, and $\Gamma t = N = 0$.}\label{f:fidNoise}
\end{figure}

When the shared state is the TWB of Eq.~(\ref{twb:wig}), the average
fidelity is obtained from Eq.~(\ref{gen:fid}) with $F=G=2\widetilde{A}_0$
and $H=2\widetilde{B}_0$, i.e.,
\begin{equation}
\overline{F}_{\hbox{\tiny TWB}}(\lambda) = \mbox{$\frac12$}(1+\lambda)
\end{equation}
whereas in the presence of noise one should use the substitutions
(\ref{sostituzioni}). $\overline{F}_{\hbox{\tiny TWB}}$ is plotted in
Fig.~\ref{f:fid:TWB}.
When the teleportation is assisted by IPS, then the fidelity reads as
follows:
\begin{equation}
\overline{F}_{\hbox{\tiny IPS}}
= \frac{1}{p_{11}(r,T,\varepsilon)} \sum_{k=1}^{4}
\frac{{\cal C}_k}{F_k G_k-H_k^2+F_k+G_k-2H_k}\,,
\end{equation}
with $F_k = b-f_h$, $G_k = b-g_h$, and $H_k = 2 \witB_0 T + h_k$ and all
the involved quantities are the same as in Eq.~(\ref{ips:wigner}). The
results are presented in Fig.~\ref{f:fidid} for $\varepsilon = 1$ and
$\Gamma t = N =0$.  The IPS state improves the average fidelity of quantum
teleportation when $\lambda$ is below a certain threshold, which depends on
$T$ (and $\varepsilon$).  Notice that, for $T< 0.5$,
$\overline{F}_{\hbox{\tiny IPS}}(\lambda)$ is always below
$\overline{F}_{\hbox{\tiny TWB}}(\lambda)$, at least for $\varepsilon=1$.
The effect of dissipation and thermal noise is shown in
Fig.~\ref{f:fidNoise}.
\par
\begin{figure}[tb]
\begin{center}
\includegraphics{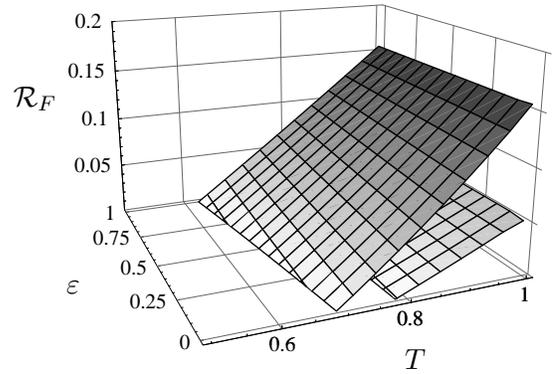}
\end{center}
\vspace{-0.5cm}
\caption{Plots of ${\cal R}_{F}$ as a function of $T$ and $\varepsilon$
with $r = 0.3$; we set $\Gamma t = N = 0$ (upper surface) and $\Gamma t = N
= 0.1$ (lower surface). ${\cal R}_{F}$, and, in turn,
$\overline{F}_{\mbox{\tiny IPS}}$, is mainly affected by the IPS
transmissivity $T$.}\label{f:fid3D}
\end{figure}
\begin{figure}[tb]
\begin{center}
\includegraphics{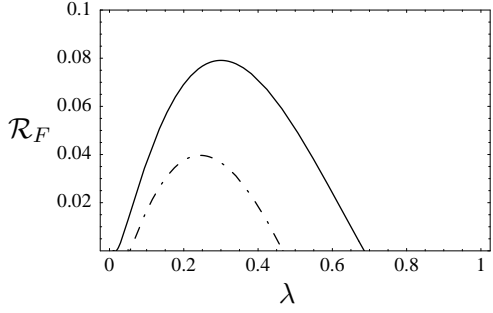}
\end{center}
\vspace{-0.5cm}
\caption{Plot of the relative enhancement ${\cal R}_{F}$ as a function of
$\lambda = \tanh r$ with $T = 0.9$, $\varepsilon = 1$, and $\Gamma t = N
=0.1$ (solid line).  The dot-dashed line is ${\cal R}_{F}^{\rm (id)}$,
namely, the relative enhancement of the fidelity using the de-Gaussified
TWB in noisy environment with respect to the fidelity using TWB in ideal
case (see text for details): for a suitable choice of the parameters, the
teleportation assisted by IPS in the presence of dissipation and thermal
noise, can have a fidelity larger than the one of TWB-assisted
teleportation also when this is implemented in ideal conditions (i.e.,
$\Gamma t = N = 0$).}\label{f:fidperN}
\end{figure}

In order to quantify the improvement and to study its dependence on $T$ and
$\varepsilon$, we define the following ``relative improvement'':
\begin{equation}
{\cal R}_{F}(r,T,\varepsilon,\Gamma,N) =
\frac{\overline{F}_{\hbox{\tiny IPS}}(r,T,\varepsilon,\Gamma,N)-
\overline{F}_{\hbox{\tiny TWB}}(r,\Gamma,N)}
{\overline{F}_{\hbox{\tiny TWB}}(r,\Gamma,N)}\,,
\end{equation}
which is plotted in Fig.~\ref{f:fid3D}: we can see that ${\cal R}_{F}$ and,
in turn, $\overline{F}_{\hbox{\tiny IPS}}$ are mainly affected by $T$ when
$\Gamma t$ and $N$ are fixed.
In Fig.~\ref{f:fidperN} we plot ${\cal R}_{F}$ as a function $\lambda=\tanh
r$ and the quantity ${\cal R}_{F}^{\rm (id)}$ defined as follows:
\begin{equation}
{\cal R}_{F}^{\rm (id)}(r,T,\varepsilon,\Gamma,N)=
\frac{\overline{F}_{\hbox{\tiny IPS}}(r,T,\varepsilon,\Gamma,N)-
\overline{F}_{\hbox{\tiny TWB}}(r,0,0)}
{\overline{F}_{\hbox{\tiny TWB}}(r,0,0)}\,,
\end{equation}
i.e., the relative improvement of the fidelity using IPS in the presence of
losses and thermal noise with respect to the fidelity using the TWB in
ideal conditions ($\Gamma t = N = 0$): we can see that, for the particular
choice of the parameters, not only the fidelity is improved with respect
the TWB-based teleportation in the presence of the same dissipation and
thermal noise (solid line in Fig.~\ref{f:fidperN}), but the results can be
also better than the ideal case (dot-dashed line). We can conclude that IPS
onto TWB degraded by dissipation and noisy environment can improve the
fidelity of teleportation up to and beyond the value achievable using the
TWB in ideal conditions.
\begin{figure}[tb]
\begin{center}
\includegraphics{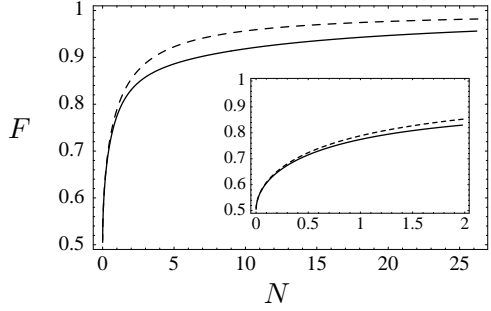}
\end{center}
\vspace{-0.5cm}
\caption{Plot of the teleportation fidelity as a function of the average
number of photons $N$ of the shared state in the case of TWB (dashed line)
and a photon subtracted TWB (solid line) for $T = 0.999$, $\varepsilon =
1$, and in ideal conditions (i.e., $\Gamma t = N = 0$. The inset is a
magnification of the region $0<N<2$.}\label{f:fid:energy}
\end{figure}
\par
Finally, in Fig.~\ref{f:fid:energy} we plot the teleportation fidelity as a
function of the average number of photons $N$ of the shared state in the
case of TWB and a photon subtracted TWB: we can see that for a fixed energy
of the shared quantum channel the best fidelity is achieved by the TWB
state. The same result holds in the presence of dissipation and thermal
noise.
\par
In the next Sections we will analyze the nonlocality of the IPS state in
the presence of noise by means of Bell's inequalities \cite{OP:PSnoise}.
\section{Nonlocality in the phase space} \label{s:DP}
Parity is a dichotomic variable and thus can be used to establish
Bell-like inequalities \cite{CHSH}. 
The displaced parity operator on two modes is defined as \cite{bana}
\begin{equation}
\hat{\Pi}(\alpha,\beta) =
D_a(\alpha)(-1)^{a^\dag a}D_a^\dag(\alpha)
\otimes D_b(\beta)(-1)^{b^\dag b}D_b^\dag(\beta)\,,
\end{equation}
where $\alpha, \beta \in {\mathbb C}$, $a$ and $b$ are mode operators and
$D_a(\alpha)=\exp\{\alpha a^\dag - \calpha a\}$ and $D_b(\beta)$ are
single-mode displacement operators.  Since the two-mode Wigner function
$W(\alpha,\beta)$ can be expressed as \cite{FOP:napoli:05}
\begin{equation}
W(\alpha,\beta) = \frac{4}{\pi^2}\, \Pi(\alpha,\beta)\,,
\end{equation}
$\Pi(\alpha,\beta)$ being the expectation value of $\hat\Pi(\alpha,\beta)$,
the violation of these inequalities is also known as nonlocality in the
phase-space. The quantity involved in such inequalities can be written as
follows
\begin{equation}\label{bell:general}
{\cal B}_{\rm DP} = \Pi(\alpha_1,\beta_1)+ \Pi(\alpha_2,\beta_1)
+ \Pi(\alpha_1,\beta_2)-\Pi(\alpha_2,\beta_2)\,,
\end{equation}
which, for local theories, satisfies $|\mathcal{B}_{\rm DP}|\le 2$.
\par
Following Ref.~\cite{bana}, one can choose a particular set of
displaced parity operators, arriving at the following combination
\cite{ips:PRA:70}
\begin{multline}
{\cal B}_{\rm DP}({\cal J}) =
\Pi(\sqrt{\cal J},-\sqrt{\cal J})+ \Pi(-3\sqrt{\cal J},-\sqrt{\cal J})\\
+ \Pi(\sqrt{\cal J},3\sqrt{\cal J})-\Pi(-3\sqrt{J},3\sqrt{\cal J})\,,
\label{bell:ale}
\end{multline}
which, for the TWB, gives a maximum ${\cal B}_{\rm DP} = 2.32$ (for ${\cal
J} = 1.6 \times 10^{-3}$) greater than the value $2.19$ obtained in
Ref.~\cite{bana}. Notice that, even in the infinite squeezing limit, the
violation is never maximal, i.e., $|\mathcal{B}_{\rm DP}| < 2\sqrt{2}$
\cite{jeong1}.
\par
In Ref.~\cite{ips:PRA:70} we studied Eq.~(\ref{bell:ale}) for both the TWB
and the IPS state in an ideal scenario, namely in the absence of
dissipation and noise; we showed that, using IPS, the maximum violation
is achieved for $T,\varepsilon \to 1$ and for values of $r$ smaller than
for the TWB.
\par
\begin{figure}
\setlength{\unitlength}{1mm}
\begin{center}
\includegraphics[width=7cm]{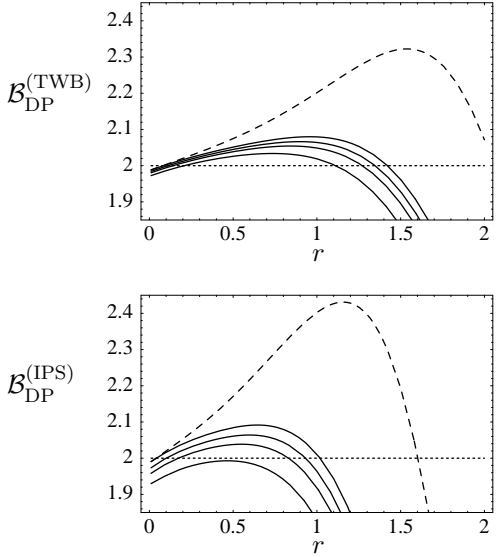}
\end{center}
\vspace{-.5cm}
\caption{Plots of the Bell parameters ${\cal B}_{\rm DP}$ for the TWB (top)
and IPS (bottom); we set ${\cal J}=1.6 \times 10^{-3}$, which maximizes
${\cal B}_{\rm DP}^{\rm (TWB)}$, and put $T = 0.9999$ and $\varepsilon
= 1$ for the IPS.  The dashed lines refer to the absence of noise ($\Gamma
t = N = 0$), whereas, for both the plot, the solid lines are ${\cal B}_{\rm
DP}$ with $\Gamma t = 0.01$ and, from top to bottom, $N=0, 0.05, 0.1,$ and
$0.2$. In the ideal case the maxima are ${\cal B}_{\rm DP}^{\rm
(TWB)}=2.32$ and ${\cal B}_{\rm DP}^{\rm (IPS)}=2.43$, respectively.}
\label{f:DP}
\end{figure}
Now, by means of the Eq.~(\ref{ips:wigner}) and the substitutions
(\ref{sostituzioni}), we can study how noise affects ${\cal B}_{\rm DP}$.
The results are showed in Fig.~\ref{f:DP} for $\varepsilon = 1$: as one may
expect, the overall effect of noise is to reduce the violation of the
Bell's inequality. When dissipation alone is present ($N=0$), the maximum
of violation is achieved using the IPS for values of $r$ smaller than for
the TWB, as in the ideal case. On the other hand, one can see that the
presence of thermal noise mainly affects the IPS results. In fact, for
$\Gamma t = 0.01$ and $N=0.2$, one has $|{\cal B}_{\rm DP}^{\rm (TWB)}|>2$
for a range of $r$ values, whereas $|{\cal B}_{\rm DP}^{\rm (IPS)}|$ falls
below the threshold for violation. Note that the maximum of violation, both
for the TWB and the IPS state, depends on the squeezing parameter $r$.
\par
\begin{figure}[t!]
\vspace{-0.5cm}
\setlength{\unitlength}{1mm}
\begin{center}
\begin{picture}(70,60)(-10,0)
\put(0,0){\includegraphics[width=6cm]{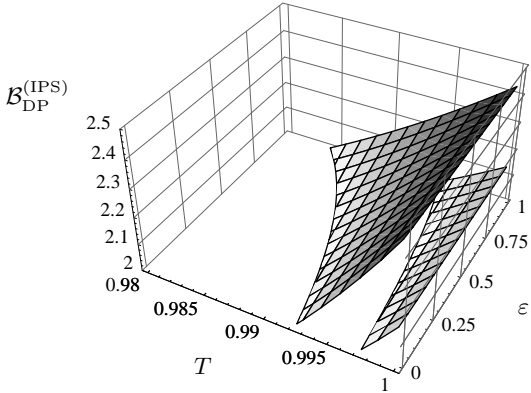}}
\put(58,12){$\varepsilon$}
\put(15,4){$T$}
\put(-10,40){${\cal B}_{\rm DP}^{\rm (IPS)}$}
\end{picture}
\end{center}
\vspace{-0.5cm}
\caption{The surfaces are plots of the Bell parameters
${\cal B}_{\rm DP}$ for the IPS state as a function of $T$ and 
$\varepsilon$ for different values of $\Gamma t$ and $N = 0$:
(top) $\Gamma t = 0$; (bottom) $\Gamma t = 0.005$.
We set ${\cal J}=1.6 \times 10^{-3}$ and $r = 1.16$
The value of the Bell parameter is mainly affected by $T$.}
\label{f:DP_3D}
\end{figure}
In Fig.~\ref{f:DP_3D} we plot ${\cal B}_{\rm DP}^{\rm (IPS)}$ as a function
of $T$ and $\varepsilon$. We can see that the main contribution to the
Bell parameter is due to the transmissivity $T$. Moreover, as $T \to
1$, the Bell parameter is actually independent on $\varepsilon$.
Note that the values of ${\cal J}$ and $r$, which maximize the violation,
depend on $\Gamma t$ and $N$, as one can see from Fig.~\ref{f:DP}: in
Fig.~\ref{f:DP_3D} we have chosen to fix the environmental parameters in
order to compare the two plots, even if best results can be obtained
maximizing ${\cal B}_{\rm DP}^{\rm (IPS)}$ with respect ${\cal J}$ and
$T$.
\par
We conclude that, considering the displaced parity test in the presence
of noise, the IPS is quite robust if the thermal noise is below a threshold
value (depending on the environmental parameters) and for small values of the
TWB parameter $r$.
\section{Nonlocality and homodyne detection} \label{s:HD}
In principle there are two approaches how to test the Bell's inequalities
for bipartite state: either one can employ some test for continuous variable
systems, such as that described in Sec.~\ref{s:DP}, or one can convert the
problem to Bell's inequalities tests on two qubits by mapping the
two modes into two-qubit systems. In this and the following Section we
will consider this latter case.
\par
The Wigner function $W_{\hbox{\tiny IPS}}(\alpha,\beta)$ given in
Eq.~(\ref{ips:wigner}) is no longer positive-definite and thus
it can be  used to test the violation of Bell's
inequalities by means of  homodyne detection, i.e., measuring the
quadratures $x_{\vartheta}$ and $x_{\varphi}$ of the two IPS modes $a$ and
$b$, respectively, as proposed in Refs.~\cite{nha,sanchez}.
In this case, one can dichotomize the measured quadratures assuming as
outcome $+1$ when $x \ge 0$, and $-1$ otherwise. The nonlocality of
$W_{\hbox{\tiny IPS}}(\alpha,\beta)$ in ideal conditions has been studied in
Ref.~\cite{ips:PRA:70} where we also discussed the effect of the homodyne
detection efficiency $\eta_{\rm H}$.
\par
Let us now we focus our attention on $W_{\hbox{\tiny IPS}}(\alpha,\beta)$
when the IPS process is applied to the TWB evolved through the noisy
channel, namely, using the substitutions (\ref{sostituzioni}).  After the
dichotomization of the homodyne outputs, one obtains the following Bell
parameter
\begin{equation}\label{bell:homo}
{\cal B}_{\rm HD} =
E(\vartheta_1,\varphi_1) + E(\vartheta_1,\varphi_2)
+ E(\vartheta_2,\varphi_1) - E(\vartheta_2,\varphi_2)\,,
\end{equation}
where $\vartheta_k$ and $\varphi_k$ are the phases of the two
homodyne measurements at the modes $a$ and $b$, respectively, and
\begin{equation}
E(\vartheta_h,\varphi_k) =
\int_{\mathbb{R}^2} d x_{\vartheta_h}\,d x_{\varphi_k}\,
{\rm sign}[x_{\vartheta_h}\, x_{\varphi_k}]\,
P(x_{\vartheta_h}, x_{\varphi_k})\,,
\end{equation}
$P(x_{\vartheta_h}, x_{\varphi_k})$ being the joint
probability of obtaining the two outcomes
$x_{\vartheta_h}$ and $x_{\varphi_k}$ \cite{sanchez}. As usual,
violation of Bell's inequality is achieved when $|{\cal B}_{\rm HD}|>2$.
\par
\begin{figure}[tb]
\vspace{-1cm}
\setlength{\unitlength}{1mm}
\begin{center}
\begin{picture}(70,100)(0,0)
\put(4,0){\includegraphics[width=6.5cm]{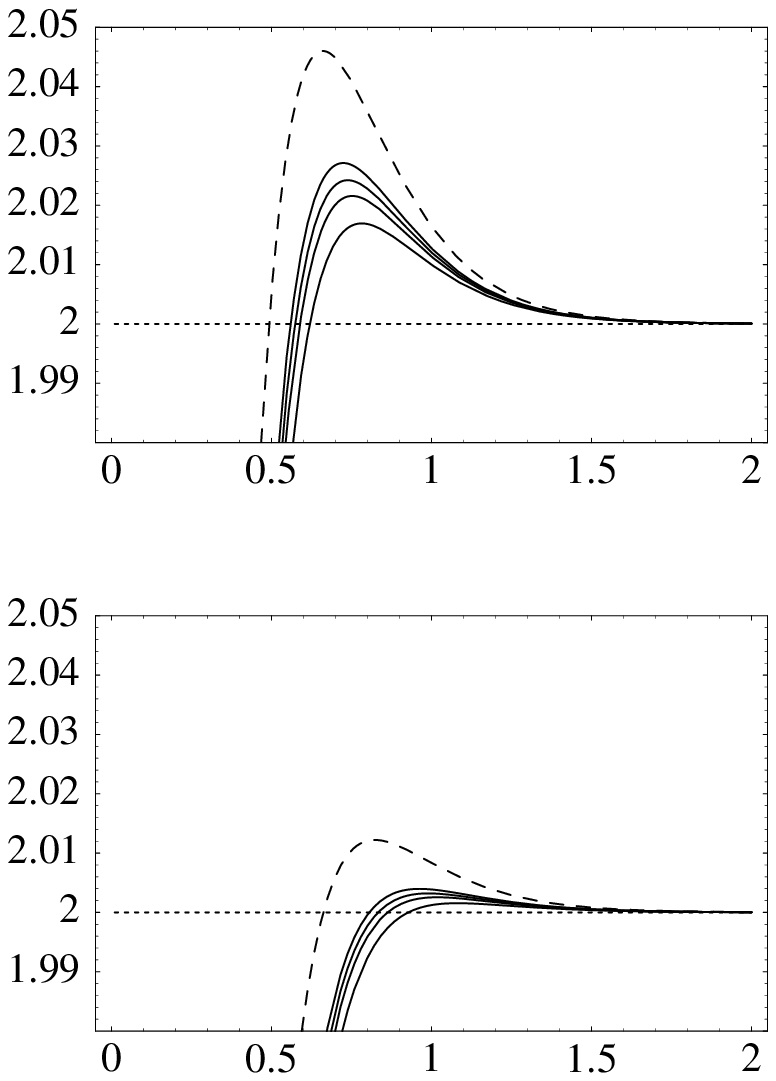}}
\put(37,45){$r$}
\put(-2,70){${\cal B}_{\rm HD}$}
\put(37,-1){$r$}
\put(-2,24.5){${\cal B}_{\rm HD}$}
\end{picture}
\end{center}
\caption{Plots of the Bell parameter ${\cal B}_{\rm HD}$ for the IPS states
for two different values of the homodyne detection efficiency: $\eta_{\rm
H} = 1$ (top), and $\eta_{\rm H}=0.9$ (bottom). We set $\varepsilon = 1$
and $T = 0.99$. The dashed lines refer to the absence of noise ($\Gamma
t = N = 0$), whereas, for both the plots, the solid lines are ${\cal
B}_{\rm HD}$ with $\Gamma t = 0.05$ and, from top to bottom, $N=0, 0.05,
0.1$ and $0.2$.} \label{f:HD}
\end{figure}
In Fig.~\ref{f:HD} we plot ${\cal B}_{\rm HD}$ for $\vartheta_1 = 0$,
$\vartheta_2 = \pi/2$, $\varphi_1 = -\pi/4$ and $\varphi_2 = \pi/4$: as for
the ideal case \cite{ips:PRA:70,sanchez}, the Bell's inequality is
violated for a suitable choice of the squeezing parameter $r$. Obviously,
the presence of noise reduces the violation, but we can see that the effect
of thermal noise is not so large as in the case of the displaced parity
test addressed in Sec.~\ref{s:DP} (see Fig.~\ref{f:DP}).
\begin{figure}[t!]
\vspace{-0.5cm}
\setlength{\unitlength}{1mm}
\begin{center}
\begin{picture}(70,60)(-10,0)
\put(0,0){\includegraphics[width=6cm]{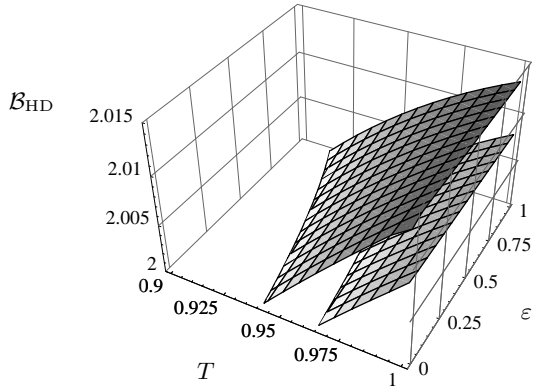}}
\put(58,12){$\varepsilon$}
\put(15,4){$T$}
\put(-10,40){${\cal B}_{\rm HD}$}
\end{picture}
\end{center}
\vspace{-0.5cm}
\caption{The surfaces are plots of the Bell parameters
${\cal B}_{\rm HD}$ for the IPS state as a function of $T$ and 
$\varepsilon$ for $N = 0$ and different values of $\Gamma t$:
(top) $\Gamma t = 0$, and (bottom) $\Gamma t = 0.025$. We set $r
= 0.82$ and $\eta_{\rm H} = 0.9$}
\label{f:HD_3D}
\end{figure}
In Fig.~\ref{f:HD_3D} we plot ${\cal B}_{\rm HD}$ as a function of $T$
and $\varepsilon$: as for the displaced parity test (see
Fig.~\ref{f:DP_3D}), we can see that the main contribution to the
Bell parameter is due to the transmissivity $T$.
\par
Notice that the high efficiencies of this kind of detectors
allow a loophole-free test of hidden variable theories
\cite{gil}, though the violations obtained are quite small.
This is due to the intrinsic information loss of the binning
process, which is used to convert the continuous homodyne data in
dichotomic results \cite{mun1}.
\section{Nonlocality and pseudospin test} \label{s:PS}
Another way to map a two-mode continuous variable system into a two-qubit
system is by means of the pseudospin test: this consists in measuring
three single-mode Hermitian operator $S_k$ satisfying the Pauli matrix algebra
$[S_h,S_k]=2i\varepsilon_{hkl}\,S_l$, $S_k^2 = {\mathbb I}$, $h,k,l=1,2,3$,
and $\varepsilon_{hkl}$ is the totally antisymmetric tensor with
$\varepsilon_{123}=+1$ \cite{filip:PRA:66,chen:PRL:88}. For the sake of
clarity, we will refer to $S_1$, $S_2$ and $S_3$ as $S_x$, $S_y$ and $S_z$,
respectively. In this way one can write the following correlation function
\begin{equation}
E({\bf a},{\bf b}) = \langle ({\bf a}\cdot{\bf S})\,
({\bf b}\cdot{\bf S})\rangle\,,
\end{equation}
where ${\bf a}$ and ${\bf b}$ are unit vectors such that
\begin{align}
{\bf a}\cdot{\bf S} &= \cos \vartheta_a\, S_z +
\sin \vartheta_a\, (e^{i \varphi_a} S_{-} + e^{-i \varphi_a} S_{+})\,,\\
{\bf b}\cdot{\bf S} &= \cos \vartheta_b\, S_z +
\sin \vartheta_b\, (e^{i \varphi_b} S_{-} + e^{-i \varphi_b} S_{+})\,,
\end{align}
with $S_{\pm} = \frac12 (S_x \pm i S_y)$. In the following, without loss of
generality, we set $\varphi_k = 0$. Finally, the Bell parameter reads
\begin{equation}\label{bell:PS}
{\cal B}_{\rm PS} = E({\bf a}_1,{\bf b}_1)+E({\bf a}_1,{\bf b}_2)
+E({\bf a}_2,{\bf b}_1)-E({\bf a}_2,{\bf b}_2)\,,
\end{equation}
corresponding to the CHSH Bell's inequality $|{\cal B}_{\rm PS}|\le 2$. In
order to study Eq.~(\ref{bell:PS}) we should choose a specific
representation of the pseudospin operators; note that, as pointed out in
Refs.~\cite{revzen, ferraro:3:nonloc}, the violation of Bell inequalities
for continuous variable systems depends, besides on the orientational
parameters, on the chosen representation, since different $S_k$ leads to
different expectation values of ${\cal B}_{\rm PS}$. Here we consider the
pseudospin operators corresponding to the Wigner functions \cite{revzen}
\begin{align}
W_x(\alpha)&=\frac{1}{\pi}\,{\rm sign}\big[\Re{\rm e}[\alpha]\big]\,,\quad
W_z(\alpha)= -\frac{1}{2}\,\delta^{(2)}(\alpha)\,,\label{PS:W:xz}\\
&W_y(\alpha)=-\frac{1}{2\pi}\, \delta\big(\Re{\rm e}[\alpha] \big)\,
{\cal P} \frac{1}{\Im{\rm m}[\alpha]}\,,
\end{align}
where ${\cal P}$ denotes the Cauchy's principal value. Thanks to
(\ref{PS:W:xz}) one obtains
\begin{multline}
E_{\rm TWB}({\bf a},{\bf b}) = \cos\vartheta_a \cos\vartheta_b \\
+ \frac{2\sin\vartheta_a \sin\vartheta_b}{\pi}\,
\arctan\big[ \sinh(2r) \big]\,,
\end{multline}
for the TWB, and, for the IPS,
\begin{multline}
E_{\rm IPS}({\bf a},{\bf b}) = 
\sum_{k=1}^4 \frac{{\cal C}_k}{p_{11}(r,T,\varepsilon)}
\Bigg[
\frac{\cos\vartheta_a \cos\vartheta_b}{4} \\
+ \frac{2 \sin\vartheta_a \sin\vartheta_b}{\pi{\cal A}_k}\,
\arctan\left( \frac{2 \wtB_0 T + h_k}{\sqrt{{\cal A}_k}} \right)
\Bigg]
\end{multline}
where $ {\cal A}_k=(b-f_k)(b-g_k)-(2 \wtB_0 T + h_k)^2$,
and all the other quantities have been defined in Sec.~\ref{s:IPS}.
\par
\begin{figure}[tb]
\vspace{-1cm}
\setlength{\unitlength}{1mm}
\begin{center}
\begin{picture}(70,50)(0,0)
\put(4,0){\includegraphics[width=6.5cm]{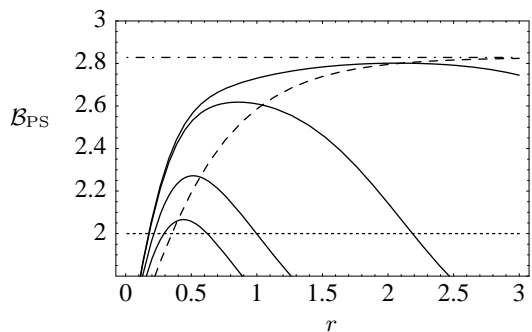}}
\put(40,-3){$r$}
\put(-2,24.5){${\cal B}_{\rm PS}$}
\end{picture}
\end{center}
\caption{Plots of the Bell parameter ${\cal B}_{\rm PS}$ in ideal case
($\Gamma t = N = 0$): the dashed line refers to the TWB, whereas the solid
lines refer to the IPS with $\varepsilon = 1$ and, from top to bottom,
$T = 0.9999, 0.99, 0.9$, and $0.8$. There is a threshold value for $r$
below which IPS gives a higher violation than TWB.  Note that there is also
a region of small values of $r$ for which the IPS state violates the Bell's
inequality while the TWB does not. The dash dotted line is the maximal
violation value $2\sqrt{2}$.} \label{f:PS:id}
\end{figure}
In Fig.~\ref{f:PS:id} we plot ${\cal B}_{\rm PS}$ for the TWB and IPS in
the ideal case, namely in the absence of dissipation and thermal noise. For
all the Figures we set $\vartheta_{a_1}=0$, $\vartheta_{a_2}=\pi/2$, and
$\vartheta_{b_1}=-\vartheta_{b_2}=\pi/4$. As
usual the IPS leads to better results for small values of $r$. Whereas
${\cal B}_{\rm PS}^{\rm (TWB)} \to 2\sqrt{2}$ as $r\to \infty$,
${\cal B}_{\rm PS}^{\rm (IPS)}$ has a maximum and, then, falls below the
threshold $2$ as $r$ increases. It is interesting to note that there is a
region of small values of $r$ for which  ${\cal B}_{\rm PS}^{\rm (TWB)}\le
2 < {\cal B}_{\rm PS}^{\rm (IPS)}$, i.e., the IPS process can increases
the nonlocal properties of a TWB which does not violates the Bell's
inequality for the pseudospin test, in such a way that the resulting state
violates it. This fact is also present in the case of the displaced parity
test described in Sec.~\ref{s:DP}, but using the pseudospin test the effect
is enhanced. Notice that the maximum violations for the IPS occur for a
range of values $r$ experimentally achievable.
\par
\begin{figure}[tb]
\vspace{-1cm}
\setlength{\unitlength}{1mm}
\begin{center}
\begin{picture}(70,50)(0,0)
\put(4,0){\includegraphics[width=6.5cm]{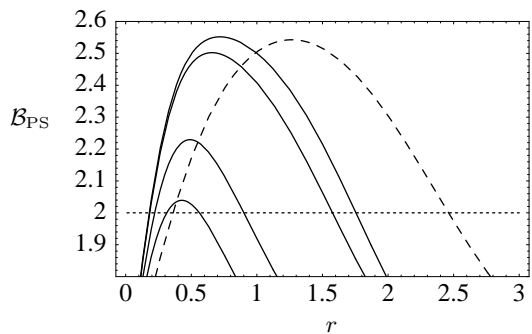}}
\put(40,-3){$r$}
\put(-2,24.5){${\cal B}_{\rm PS}$}
\end{picture}
\end{center}
\caption{Plots of the Bell parameter ${\cal B}_{\rm PS}$ for $\Gamma t =
0.01$: the dashed line refers to the TWB, whereas the solid lines refer to
the IPS with $\varepsilon = 1$ and, from top to bottom, $T = 0.9999,
0.99, 0.9$, and $0.8$.  The same comments as in Fig.~\ref{f:PS:id} still
hold.} \label{f:PS:tau}
\end{figure}
\begin{figure}[tb]
\vspace{-1cm}
\setlength{\unitlength}{1mm}
\begin{center}
\begin{picture}(70,50)(0,0)
\put(4,0){\includegraphics[width=6.5cm]{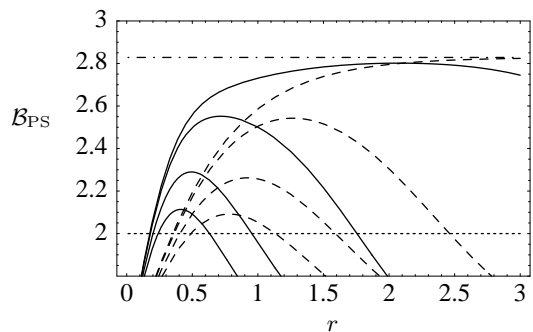}}
\put(40,-3){$r$}
\put(-2,24.5){${\cal B}_{\rm PS}$}
\end{picture}
\end{center}
\caption{Plots of the Bell parameter ${\cal B}_{\rm PS}$ for different
values of $\Gamma t$ and in the absence of thermal noise ($N = 0$): the
dashed lines refer to the TWB, whereas the solid ones refer to the IPS with
$T = 0.9999$ and $\varepsilon = 1$;
for both the TWB and IPS we set, from top to bottom,
$\Gamma t = 0, 0.01, 0.05$, and $0.1$. The dash dotted line is the maximal
violation value $2\sqrt{2}$.} \label{f:PS:gamma}
\end{figure}
\begin{figure}[tb]
\vspace{-1cm}
\setlength{\unitlength}{1mm}
\begin{center}
\begin{picture}(70,50)(0,0)
\put(4,0){\includegraphics[width=6.5cm]{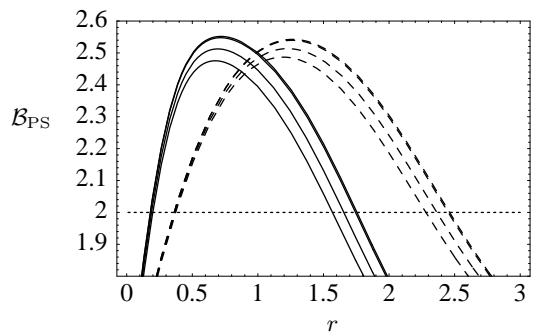}}
\put(40,-3){$r$}
\put(-2,24.5){${\cal B}_{\rm PS}$}
\end{picture}
\end{center}
\caption{Plots of the Bell parameter ${\cal B}_{\rm PS}$ for $\Gamma t =
0.01$ and different values $N = 0$: the dashed lines refer to the TWB,
whereas the solid ones refer to the IPS with $T = 0.9999$ and
$\varepsilon = 1$; for both the TWB and IPS we set, from top to bottom, $N
= 0, 0.01, 0.1$, and $0.2$.} \label{f:PS:th}
\end{figure}

In Fig.~\ref{f:PS:tau} we consider the presence of the dissipation alone
and vary $T$. We can see that IPS is effective also when the
effective transmissivity $T$ is not very high.
We take into account the effect of dissipation and thermal noise
in Figs.~\ref{f:PS:gamma}, and \ref{f:PS:th}: we can conclude that
IPS is quite robust with respect to this sources of noise and, moreover,
one can think of employing IPS as a useful resource in order to reduce the
effect of noise.
\begin{figure}[t!]
\vspace{-0.5cm}
\setlength{\unitlength}{1mm}
\begin{center}
\begin{picture}(70,60)(-10,0)
\put(0,0){\includegraphics[width=6cm]{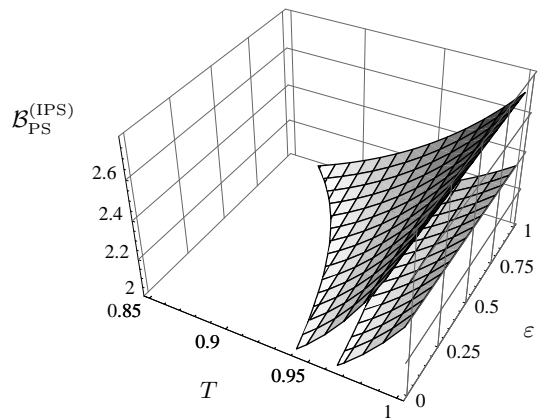}}
\put(58,12){$\varepsilon$}
\put(15,4){$T$}
\put(-10,40){${\cal B}_{\rm PS}^{\rm (IPS)}$}
\end{picture}
\end{center}
\vspace{-0.5cm}
\caption{The surfaces are plots of the Bell parameters
${\cal B}_{\rm PS}^{\rm (IPS)}$ for the IPS state as a function of
$T$ and  $\varepsilon$ for $N = 0$ and different values of $\Gamma t$:
(top) $\Gamma t = 0$, and (bottom) $\Gamma t = 0.025$. We set $r
= 0.86$.} \label{f:PS_3D}
\end{figure}
In Fig.~\ref{f:PS_3D} we plot ${\cal B}_{\rm PS}^{\rm (IPS)}$
as a function of $T$ and $\varepsilon$: the main effect on the Bell
parameter is due to the transmissivity $T$, as in the precious cases.
\section{Nonlocality and on/off photodetection}\label{s:on:off}
\begin{figure}[htbp]
\begin{center}
\includegraphics[width=85mm]{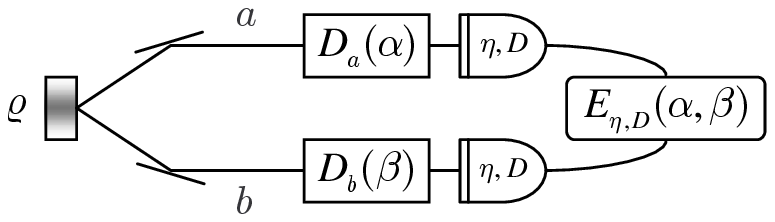}
\end{center}
\vspace{-.5cm}
\caption{Scheme of the nonlocality test based on displaced 
on/off photodetection: the two modes $a$ and $b$ of a bipartite state 
$\varrho$ are locally displaced by an amount $\alpha$ and $\beta$ 
respectively, and then revealed through on/off photodetection.
The corresponding correlation function violates Bell's
inequalities for dichotomic measurements for a suitable choice of the 
parameters $\alpha$ and $\beta$, depending on the kind of state
under investigation. The violation holds also for non-unit quantum 
efficiency and non-zero dark counts.
}\label{f:Donoff}
\end{figure}
The nonlocality test we are going to analyze is schematically depicted in
Fig.~\ref{f:Donoff}: two modes of the de-Gaussified TWB radiation field,
$a$ and $b$, described by the density matrix $\varrho$, are locally
displaced by an amount $\alpha$ and $\beta$ respectively and, finally,
they are revealed by on/off
photodetectors, i.e., detectors which have no output when no photon is
detected and a fixed output when one or more photons are detected. The
action of an on/off detector is described by the following two-value
positive operator-valued measure (POVM) $\{\Pi_{0,\eta,D},
\Pi_{1,\eta,D}\}$ \cite{FOP:napoli:05}
\begin{subequations}
\label{povm1}
\begin{align}
&{\Pi}_{0,\eta,D} = \frac1{1+D}\: \sum_{k=0}^{\infty}
\left( 1-\frac{\eta}{1+D} \right)^k \ket{k}\bra{k}\:,\\
&{\Pi}_{1,\eta,D} = \mathbb{I} - {\Pi}_{0,\eta,D}\:,\end{align}
\end{subequations}
$\eta$ being the quantum efficiency and $D$ the mean number 
of dark counts, i.e., of clicks with vacuum input. 
In writing Eq.~(\ref{povm1}) we have considered a thermal background
as the origin of dark counts. An analogous expression may be written
for a Poissonian background \cite{IOP:05}. For small 
values of the mean number $D$ of dark counts (as it generally happens at 
optical frequencies) the two kinds of background are indistinguishable.
\par
Overall, taking into account the displacement, the measurement 
on both modes $a$ and $b$ is described by the POVM (we are 
assuming the same quantum efficiency and dark counts for both the
photodetectors)
\begin{equation}
{\Pi}_{hk}^{(\eta,D)} (\alpha,\beta)=
{\Pi}_{h}^{(\eta,D)}(\alpha)\,
\otimes
{\Pi}_{k}^{(\eta,D)}(\beta)\,
\label{povmhk}\;,
\end{equation}
where $h,k = 0,1$, and ${\Pi}_{h}^{(\eta,D)}(z)\equiv  D(z)\,
{\Pi}_{h,\eta,D}\,D^{\dag}(z)$, $D(z)=\exp\left\{z a^\dag - z^* a\right\}$ 
being the displacement operator and $z\in {\mathbb C}$ a complex
parameter.
\par
In order to analyze the nonlocality of the state $\varrho$,
we introduce the following correlation function:
\begin{align}
E_{\eta,D}(\alpha,\beta) &=
\sum_{h,k=0}^{1} (-)^{h+k}\, 
\left\langle {\Pi}_{hk}^{(\eta,D)} (\alpha,\beta) 
\right\rangle \label{E:eta} \\
&= 1 + 4\, {\cal I}_{\eta,D}(\alpha,\beta) - 
2\, \left[ {\cal G}_{\eta,D}(\alpha) + {\cal Y}_{\eta,D}(\beta) \right]\:,
\nonumber
\end{align}
where
\begin{align}
&{\cal I}_{\eta,D}(\alpha,\beta) = 
\left\langle {\Pi}_{00}^{(\eta,D)} (\alpha,\beta) \right\rangle
\label{DefFunsa} \\
&{\cal G}_{\eta,D}(\alpha) = 
\left\langle {\Pi}_{0}^{(\eta,D)}(\alpha)\otimes\mathbb{I} \right\rangle 
\label{DefFunsb} \\ 
&{\cal Y}_{\eta,D}(\beta) = 
\left\langle \mathbb{I}\otimes{\Pi}_{0}^{(\eta,D)}(\beta) \right\rangle\:,
\label{DefFunsc}
\end{align}
and where $\media{ A } \equiv {\rm Tr}[\varrho\, A]$ denotes
ensemble average on both the modes.
The so-called  Bell parameter is defined by considering four different
values of the complex displacement parameters as follows:
\begin{align}
{\cal B}_{\eta,D} &= E_{\eta,D}(\alpha,\beta) +
E_{\eta,D}(\alpha',\beta) \nonumber \\
&\hspace{1cm} + E_{\eta,D}(\alpha,\beta') - E_{\eta,D}(\alpha',\beta')\\
&= \:\: 2 + 4\left\{ {\cal I}_{\eta,D}(\alpha,\beta) 
+ {\cal I}_{\eta,D}(\alpha',\beta) +
{\cal I}_{\eta,D}(\alpha,\beta')\right. \nonumber\\
&\left.\hspace{1cm}
- {\cal I}_{\eta,D}(\alpha',\beta')
- {\cal G}_{\eta,D}(\alpha) - {\cal Y}_{\eta,D}(\beta) \right\}\,.
\label{B:param}
\end{align}
Any local theory implies that $|{\cal B}_{\eta,D}|$ satisfies the 
CHSH version of the Bell inequality, i.e., $|{\cal B}_{\eta,D}|\leq 2$ 
$\forall \alpha,\alpha',\beta, \beta'$ \cite{CHSH}, while
quantum mechanical description of the same kind of experiments does not
impose this bound.
\par
Notice that using Eqs.~(\ref{povm1}) and
(\ref{DefFunsa})--(\ref{DefFunsc}), we obtain the following scaling
properties for the functions ${\cal I}_{\eta,D}(\alpha,\beta)$, ${\cal
G}_{\eta,D}(\alpha)$ and ${\cal Y}_{\eta,D}(\beta)$
\begin{align}
\label{D2etaa}
&{\cal I}_{\eta,D}(\alpha,\beta) = \left(\frac{1}{1+D}\right)^2\,
{\cal I}_{\eta/(1+D)}(\alpha,\beta)\\
\label{D2etab}
&{\cal G}_{\eta,D}(\alpha) = \frac{1}{1+D}\,{\cal G}_{\eta/(1+D)}(\alpha)\\
\label{D2etac}
&{\cal Y}_{\eta,D}(\beta) = \frac{1}{1+D}\,{\cal Y}_{\eta/(1+D)}(\beta)
\end{align}
where ${\cal I}_{\eta} = {\cal I}_{\eta,0}$,
${\cal G}_{\eta} = {\cal G}_{\eta,0}$, and
${\cal Y}_{\eta} = {\cal Y}_{\eta,0}$.
Therefore, it will be enough to study the Bell parameter
for $D=0$, namely ${\cal B}_{\eta} = {\cal B}_{\eta,0}$, and then we can use
Eqs.~(\ref{D2etaa})--(\ref{D2etac}) to take into account the effects of
non negligible dark counts. From now on we will assume $D=0$ and suppress
the explicit dependence on $D$. Notice that using expression
(\ref{B:param}) for the Bell parameter the CHSH inequality $|{\cal
B}_{\eta,D}|\leq 2$ can be rewritten as 
\begin{align}
-1 <& \:\: 
{\cal I}_{\eta,D}(\alpha,\beta) + 
{\cal I}_{\eta,D}(\alpha',\beta) +
{\cal I}_{\eta,D}(\alpha,\beta') \nonumber \\ &-
{\cal I}_{\eta,D}(\alpha',\beta') 
- {\cal G}_{\eta,D}(\alpha) - {\cal Y}_{\eta,D}(\beta) < 0
\label{CH}\;,
\end{align}
which represents the CH version of the Bell inequality for our system \cite{CH}.
\par
In order to simplify the calculations, throughout this Section we will use
the Wigner formalism. The Wigner functions associated with the elements of
the POVM (\ref{povm1}) for $D=0$ are given by \cite{IOP:05}
\begin{align}
W[\Pi_{0,\eta}] (z) &= \frac{\Delta_\eta}{\pi \eta}\,
\exp\left\{ - \Delta_\eta\, |z|^2  \right\}\,, \\
W[\Pi_{1,\eta}] (z) &= W[\iid](z) - W[\Pi_{0,\eta}] (z) \,,
\end{align}
with $\Delta_\eta = 2 \eta / (2 - \eta)$, and $W[\iid](z)=\pi^{-1}$.
Then, noticing that for any operator $O$ one has 
\begin{equation}
W[D(\alpha)\,O \,D^{\dag}(\alpha)](z) = W[O](z - \alpha)\,,
\end{equation}
it follows that
$W[D(\alpha)\,{\Pi}_{0,\eta}\,D^{\dag}(\alpha)](z)$ is given by
\begin{equation}
W[D(\alpha)\,{\Pi}_{0,\eta}\,D^{\dag}(\alpha)](z)
= W[\Pi_{0,\eta}] (z-\alpha)\,,
\end{equation}
and therefore
\begin{align}
W[\Pi_{00}^{(\eta,0)}(\alpha,\beta)](z,w)&= 
W[\Pi_{0,\eta}] (z-\alpha)\, W[\Pi_{0,\eta}] (w-\beta) \\
W[\Pi_{0,\eta}(\alpha)\otimes\iid](z,w)&= 
W[\Pi_{0,\eta}] (z-\alpha)\: \pi^{-1} \\
W[\iid \otimes \Pi_{0,\eta}(\beta)](z,w)&= 
 \pi^{-1}\:W[\Pi_{0,\eta}] (w-\beta) 
\label{uwigs}\;.
\end{align}
Finally, thanks to the trace rule expressed in the phase space
of two modes, i.e.,
\begin{equation}
{\rm Tr}[O_1\,O_2] =
\pi^2 
\int_{\mathbb{C}^2}\!d^{2}z \:d^{2}w
\, W[O_1](z,w)\,W[O_2](z,w)\,,
\end{equation}
one can evaluate the functions ${\cal I}_{\eta}(\alpha,\beta)$, 
${\cal G}_{\eta}(\alpha)$, and ${\cal Y}_{\eta}(\beta)$, and in turn 
the Bell parameter ${\cal B}_{\eta}$ in Eq.~(\ref{B:param}),
as a sum of Gaussian integrals in the complex plane.
\par
Let us now consider the TWB (\ref{twb:wig}). Since the Wigner functions of
the TWB and of the POVM (\ref{povmhk}) are Gaussian, it is quite simple
to evaluate ${\cal I}_{\eta}(\alpha, \beta)$, ${\cal G}_{\eta}(\alpha)$, and
${\cal Y}_{\eta}(\beta)$ of the correlation function (\ref{E:eta}) and, then,
${\cal B}_{\eta}$; we have
\begin{align}
&{\cal I}_{\eta} (\alpha,\beta) = 
\frac{{\cal M}_{\eta}(r)}{\eta^2\sqrt{{\rm Det}[\bmsigma_0]}}\,\nonumber \\
&\hspace{1cm}\times \exp\big\{
- \widetilde{F}_{\eta} \, (|\alpha|^2 + |\beta|^2)
+\widetilde{H}_{\eta} \, (\alpha\beta + \calpha\cbeta)\big\}\\
&{\cal G}_{\eta}(\alpha) = {\cal Y}_{\eta}(\alpha) =
\frac{\left(2\sqrt{{\rm Det}[\bmsigma_0]}\right)^{-1} \Delta_{\eta}}
{[2(\witA_0^2 - \witB_0^2) + \witA_0 \Delta_{\eta}]}\nonumber \\
&\hspace{1cm}\times
\exp\left\{ -\frac{2 \Delta_{\eta} }
{ 2(\witA_0^2 - \witB_0^2) + \witA_0 \Delta_{\eta} }\,|\alpha|^{2} \right\} 
\end{align}
with
\begin{align}
&\widetilde{F}_{\eta}
\equiv\widetilde{F}_{\eta}(r)=
\Delta_{\eta}-(2 \witA_0 + \Delta_\eta)\,{\cal M}_{\eta}(r)\\
&\widetilde{H}_{\eta} \equiv
\widetilde{H}_{\eta}(r)=  2 \witB_0 {\cal M}_{\eta}(r)\\
&{\cal M}_{\eta}(r)
= \frac{\Delta_{\eta}^2}{4(\witA_0^2 - \witB_0^2) + 4 \witA_0 \Delta_{\eta} +
\Delta_{\eta}^2 }\,,
\end{align}
\par
\begin{figure}[htbp]
\begin{center}
\includegraphics[width=70mm]{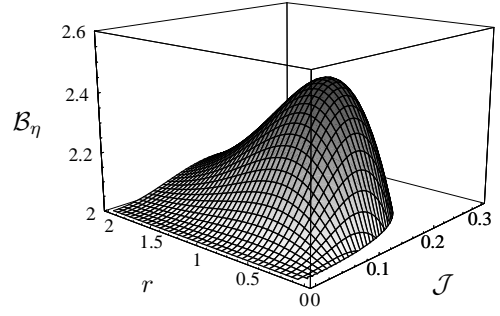}
\end{center}
\vspace{-0.5cm}
\caption{Plot of ${\cal B}_{\eta}$ for a TWB as a function of ${\cal J}$
and the TWB squeezing parameter $r$ in the case of ideal (i.e., $\eta
= 1$) on/off photodetection. The maximum violation is ${\cal
B}_{\eta}=2.45$, which is obtained when ${\cal J}=0.16$ and $r=0.74$.}
\label{f:3D}
\end{figure}
\begin{figure}[htbp]
\begin{center}
\includegraphics[width=70mm]{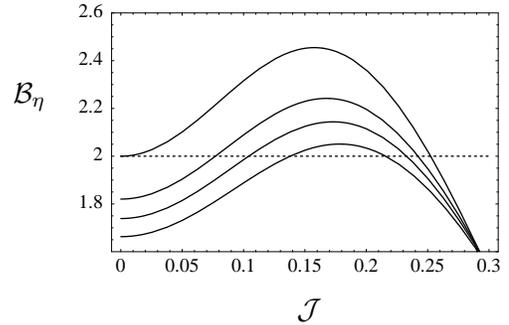}
\end{center}
\vspace{-0.5cm}
\caption{Plot of ${\cal B}_{\eta}$ for a TWB as a function of ${\cal J}$
with $r=0.74$ for different values of $\eta$: from top to bottom
$\eta=1.0$, $0.9$, $0.85$, and $0.80$.} \label{f:eta}
\end{figure}

In order to study Eq.~(\ref{B:param}), we consider the parametrization
$\alpha = -\beta = {\cal J}$ and $\alpha' = -\beta' = -\sqrt{11}{\cal J}$
(more details are given in \cite{IOP:05}).
The parametrization was chosen after a semi-analytical
analysis and maximizes the violation of the Bell's inequality (for
$\eta=1$). In Fig.~\ref{f:3D} we plot ${\cal B}_{\eta}$ for $\eta=1$:
as one can see the inequality $|{\cal B}_{\eta}|\leq 2$ is violated for a
wide range of parameters, and the maximum violation (${\cal
B}_{\eta}=2.45$) is achieved when ${\cal J}=0.16$ and $r=0.74$.
The effect of non-unit efficiency in the detection stage is to reduce the
the violation; this is shown in Fig.~\ref{f:eta}, where we plot ${\cal
B}_{\eta}$ as a function of ${\cal J}$ with $r=0.74$ for different values
of the quantum efficiency. Note that though the violation in the ideal
case, i.e., $\eta=1$, is smaller than for the Bell states, the TWBs
are more robust when one takes into account non-unit quantum
efficiency.
\par
In the case of the state (\ref{ips:wigner}), the correlation function
(\ref{E:eta}) reads (for the sake of simplicity we do not write explicitly
the dependence on $r$, $T$ and $\varepsilon$)
\begin{multline}
E_{\eta}(\alpha,\beta) = 1 +
\frac{1}{p_{11}(r,T,\varepsilon)}
\sum_{k=1}^4 {\cal C}_k\,
\big\{ 4\, {\cal I}^{(k)}_{\eta}(\alpha, \beta) \\
- 2 \big[{\cal G}^{(k)}_{\eta}(\alpha) +
 {\cal Y}^{(k)}_{\eta}(\beta)\big]\big\}\,,
\end{multline}
where
\begin{align}
&{\cal I}^{(k)}_{\eta} (\alpha,\beta) =
\frac{{\cal M}_{\eta}^{(k)}(r,T,\varepsilon)}{\eta^2}\, \nonumber\\
&\times
\exp\big\{
-\widetilde{G}_{\eta}^{(k)} \, |\alpha|^2
-\widetilde{F}_{\eta}^{(k)} \, |\beta|^2
+\widetilde{H}_{\eta}^{(k)} \, (\alpha\beta + \calpha\cbeta)
\big\}\,,\\
&{\cal G}^{(k)}_{\eta}(\alpha) =
\frac{ \Delta_{\eta}}{[G_k\,(F_k + \Delta_{\eta}) -
H_k^2]\,\eta} \nonumber\\
&\hspace{1cm}\times
\exp\left\{
-\frac{(F_k G_k - H_k^2)\,\Delta_{\eta}}{G_k\,(F_k +
\Delta_{\eta})-H_k^2}\,|\alpha|^{2} \right\}\,, \\
&{\cal Y}^{(k)}_{\eta}(\beta) =
\frac{ \Delta_{\eta}}{[F_k\,(G_k + \Delta_{\eta}) -
H_k^2]\,\eta} \nonumber\\
&\hspace{1cm}\times
\exp\left\{
-\frac{(F_k G_k - H_k^2)\,\Delta_{\eta}}{F_k\,(G_k +
\Delta_{\eta})-H_k^2}\,|\beta|^{2} \right\}\,.
\end{align}
with $\widetilde{F}_{\eta}^{(k)}
\equiv\widetilde{F}_{\eta}^{(k)}(r,T,\varepsilon)$,
$\widetilde{G}_{\eta}^{(k)}
\equiv\widetilde{G}_{\eta}^{(k)}(r,T,\varepsilon)$, and
$\widetilde{H}_{\eta}^{(k)}
\equiv\widetilde{H}_{\eta}^{(k)}(r,T,\varepsilon)$ given by
\begin{align}
&\widetilde{F}_{\eta}^{(k)} =
\Delta_{\eta}-(F_k+\Delta_\eta)\,{\cal M}_{\eta}^{(k)}(r,T,\varepsilon)\,,\\
&\widetilde{G}_{\eta}^{(k)} = 
\Delta_{\eta}-(G_k+\Delta_\eta)\,{\cal M}_{\eta}^{(k)}(r,T,\varepsilon)\,,\\
&\widetilde{H}_{\eta}^{(k)} =
H_k\,{\cal M}_{\eta}^{(k)}(r,T,\varepsilon)\,,\\
&{\cal M}_{\eta}^{(k)}(r,T,\varepsilon)
= \frac{\Delta_{\eta}^2}{(F_k+\Delta_{\eta})(G_k+\Delta_{\eta}) - H_k^2}\,,
\end{align}
where $F_k = b-f_h$, $G_k = b-g_h$, and $H_k = 2 \witB_0 T - h_k$ and all
the involved quantities are the same as in Eq.~(\ref{ips:wigner}).
\par
\begin{figure}[tb]
\vspace{1cm}
\begin{center}
\includegraphics[width=70mm]{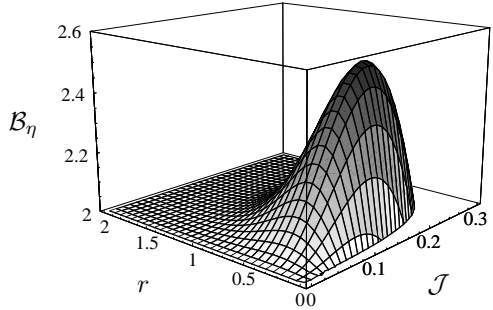}
\end{center}
\vspace{-0.3cm}
\caption{Plot of ${\cal B}_{\eta}$ for the IPS state with $T = 0.9999$ and
$\varepsilon = 1$
as a function of ${\cal J}$ and the TWB squeezing parameter $r$ in the case
of ideal (i.e., $\eta = 1$) on/off photodetection. The maximum
violation is ${\cal B}_{\eta}=2.53$, which is obtained when ${\cal J}=0.16$
and $r=0.39$.} \label{f:IPS3D}
\end{figure}
\begin{figure}[tb]
\begin{center}
\includegraphics[width=70mm]{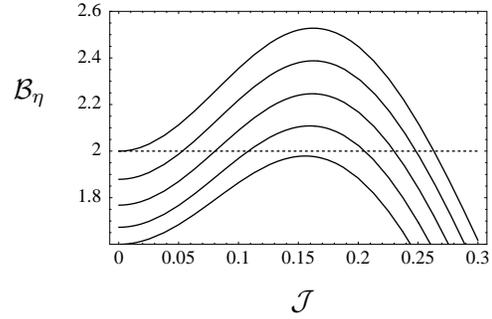}
\end{center}
\vspace{-0.5cm}
\caption{Plot of ${\cal B}_{\eta}$ for the IPS state as a function of
${\cal J}$ with $r=0.39$ for different values of $T$ and $\varepsilon = 1$
in the ideal case (i.e., $\eta = 1$): from top to bottom
$T=0.9999$, $0.90$, $0.80$, $0.70$, and $0.60$.} \label{f:IPStau}
\end{figure}

In order to study Eq.~(\ref{B:param}), we consider the parametrization
$\alpha = -\beta = {\cal J}$ and $\alpha' = -\beta' = -\sqrt{11}{\cal J}$.
This parametrization was chosen after a semi-analytical analysis and
maximizes the violation of the Bell's inequality (for $\eta=1$)
\cite{IOP:05}.  The results are showed in Figs.~\ref{f:IPS3D} and
\ref{f:IPStau} for $\eta = 1$ and $\varepsilon = 1$: we can see that the
IPS enhances the violation of the inequality $|{\cal B}_{\eta}|\le 2$ for
small values of $r$ (see also Refs.~\cite{OP:PSnoise,ips:PRA:67, ips:PRA:70}).
Moreover, as one may expect, the maximum of violation is
achieved as $T \to 1$, whereas decreasing the effective transmission of the
IPS process, one has that the inequality becomes satisfied for all the
values of $r$, as we can see in Fig.~\ref{f:IPStau} for $T = 0.6$.
\par
\begin{figure}[tb]
\begin{center}
\includegraphics[width=70mm]{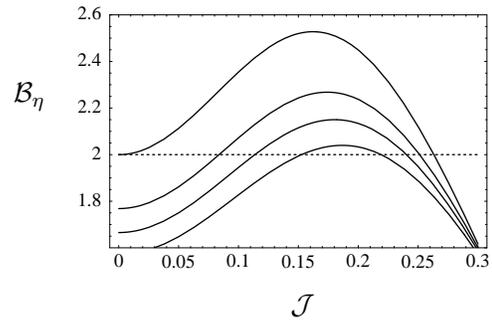}
\end{center}
\vspace{-0.5cm}
\caption{Plot of ${\cal B}_{\eta}$ for the IPS state as a function of
${\cal J}$ with $r=0.39$, $T = 0.9999$, $\varepsilon = 1$, 
and for different values of
$\eta$: from top to bottom $\eta=1.0$, $0.9$, $0.85$, and $0.8$.}
\label{f:IPSeta}
\end{figure}
\begin{figure}[tb]
\begin{center}
\includegraphics[width=70mm]{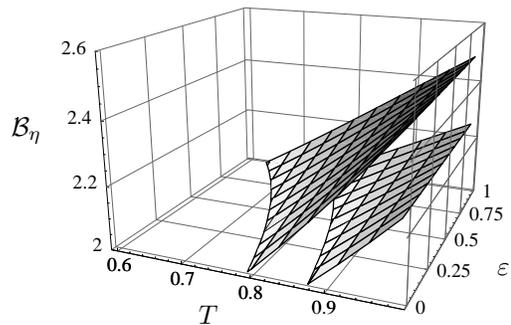}
\end{center}
\vspace{-0.5cm}
\caption{Plot of ${\cal B}_{\eta}$ for the IPS state as a function $T$ and
$\varepsilon$ with ${\cal J} = 0.16$, $r=0.39$, and, from top to bottom,
$\eta = 0.99$, and $0.90$. The main effect on ${\cal B}_{\eta}$ is due to
the transmissivity $T$.} \label{f:IPSTauEta}
\end{figure}

In Fig.~\ref{f:IPSeta} we plot ${\cal B}_{\eta}$ for the IPS
with $T = 0.9999$, $\varepsilon = 1$ and different $\eta$.
As for the TWB, we can have
violation of the Bell's inequality also for detection efficiencies near to
$80 \%$. As for the Bell states and the TWB, a $\eta$- and $r$-dependent
choice of the parameters in Eq.~(\ref{B:param}) can improve this result.
The effect on a non-unit $\varepsilon$ is studied in
Fig.~\ref{f:IPSTauEta}, where we plot ${\cal B}_\eta$ as a function of $T$
and $\varepsilon$ and fixed values of the other involved parameters.
We can see that the main effect on the Bell parameter is due to the
transmissivity $T$.
\par
\begin{figure}[tb]
\begin{center}
\includegraphics[width=70mm]{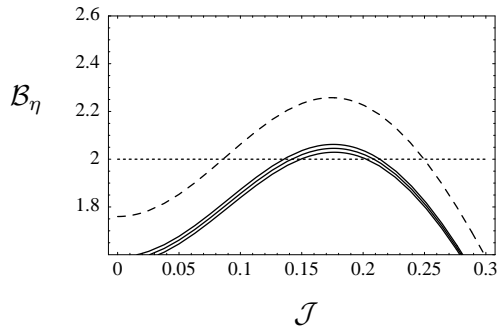}
\end{center}
\vspace{-0.5cm}
\caption{Plot of ${\cal B}_{\eta}$ for the IPS state as a function of
${\cal J}$ with $r=0.39$, $T = 0.99$, $\varepsilon = 1$, $\eta = 0.9$,
and $\Gamma t=0.1$ for different values of $N$: from top to bottom (solid
lines) $N=0$, $0.01$, and $0.02$. The dashed line is ${\cal B}_{\eta}$ with
$\Gamma t = N = 0$.}
\label{f:IPSN}
\end{figure}

Finally, the effect of dissipation and thermal noise affecting the
propagation of the TWB before the IPS process is shown in
Fig.~\ref{f:IPSN}.
\section{Conclusions} \label{s:remarks}
We have analyzed in details a photon subtraction scheme to 
de-Gaussify states of radiation and, in particular, to 
enhance nonlocal properties of twin-beams. The scheme is 
based on conditional inconclusive subtraction of photons (IPS), 
which may be achieved by means of linear optical components
and avalanche on/off photodetectors. The IPS process can 
be implemented with current technology and, indeed, application 
to single-mode state has been recently realized with high 
conditional probability \cite{weng:PRL:04}.
\par
We found that IPS process improves fidelity of 
coherent state teleportation and show, by using several 
different nonlocality tests, that it also enhances nonlocal
correlations. IPS may be profitably used also on 
nonmaximmally mixed entangled states, 
as the ones coming from the evolution of TWB in a noisy channel.
In addition, the effectiveness of the process is not dramatically
influenced either by the transmissivity 
of the beam-splitter used to subtract photons, not by 
the quantum efficiency of the detectors used to reveal them.
\par
We conclude that IPS on TWB is a robust and realistic scheme to 
improve quantum information processing with CV radiation states.
\section{Acknowledgments} 
This work has been supported by MIUR through the project 
PRIN-2005024254-002.

\end{document}